\documentclass[journal]{IEEEtran}
\usepackage{graphicx}
\usepackage{cite}
\usepackage{picinpar}
\usepackage{amsmath}
\usepackage{amsfonts}
\usepackage{url}
\usepackage{flushend}
\usepackage[latin1]{inputenc}
\usepackage{colortbl}
\usepackage{soul}
\usepackage{multirow}
\usepackage{pifont}
\usepackage{color}
\usepackage{alltt}
\usepackage[hidelinks]{hyperref}
\usepackage{enumerate}
\usepackage{siunitx}
\usepackage{breakurl}
\usepackage{epstopdf}
\usepackage{pbox}
\usepackage{mathrsfs}
\usepackage{amsfonts}
\usepackage{amssymb}
\usepackage{bm}
\usepackage{multirow}
\usepackage{booktabs}
\usepackage{algorithm}
\usepackage{algpseudocode}
\usepackage{arydshln}
\usepackage{subfigure}
\usepackage{nomencl}
\usepackage{setspace}
\makenomenclature

\def\ba{\begin{array}}
\def\ea{\end{array}}

\makeatletter
\newcommand*\bigcdot{\mathpalette\bigcdot@{.5}}
\newcommand*\bigcdot@[2]{\mathbin{\vcenter{\hbox{\scalebox{#2}{$\m@th#1\bullet$}}}}}
\makeatother

\begin{document}
\title{Detecting False Data Injection Attacks in Smart Grids with Modeling Errors: A Deep Transfer Learning Based Approach}
\author{
	Bowen~Xu,
Fanghong~Guo, \IEEEmembership{Member, IEEE},
Changyun~Wen, \IEEEmembership{Fellow, IEEE},
Ruilong~Deng, \IEEEmembership{Senior Member, IEEE},
Wen-An~Zhang, \IEEEmembership{Member, IEEE}	
\thanks{This work has been submitted to the IEEE for possible publication. Copyright may be transferred without notice, after which this version may no longer be accessible.}
\thanks{B. Xu, F. Guo and W.-A. Zhang are with the Department of Automation, Zhejiang University of Technology, Hangzhou, China. Email: bwxu@zjut.edu.cn (B. Xu), fhguo@zjut.edu.cn (F. Guo), wazhang@zjut.edu.cn (W.-A. Zhang). (\emph{Corresponding Author: F. Guo})}
\thanks{C. Wen is with the School of Electrical and Electronic Engineering, Nanyang Technological University, Singapore. Email: ecywen@ntu.edu.sg (C. Wen). }
\thanks{R. Deng is with the College of Control Science and Engineering, Zhejiang University, Hangzhou 310027, China, where he is also with the School of Cyber Science and Technology. Email: dengruilong@zju.edu.cn (R. Deng).}
}


\maketitle
\begin{abstract} Most traditional false data injection attack (FDIA) detection approaches rely on a key assumption, i.e., the power system can be accurately modeled. However, the transmission line parameters are dynamic and cannot be accurately known during operation and thus the involved modeling errors should not be neglected. In this paper, an illustrative case has revealed that modeling errors in transmission lines significantly weaken the detection effectiveness of conventional FDIA approaches. To tackle this issue, we propose an FDIA detection mechanism from the perspective of transfer learning. Specifically, the simulated power system is treated as a source domain, which provides abundant simulated normal and attack data. The real world's running system whose transmission line parameters are unknown is taken as a target domain where sufficient real normal data are collected for tracking the latest system states online. The designed transfer strategy that aims at making full use of data in hand is divided into two optimization stages. In the first stage, a deep neural network (DNN) is built by simultaneously optimizing several well-designed objective terms with both simulated data and real data, and then it is fine-tuned via real data in the second stage. Several case studies on the IEEE 14-bus and 118-bus systems verify the effectiveness of the proposed mechanism.
  

\end{abstract}


\IEEEpeerreviewmaketitle
\begin{spacing}{1}
\section{Introduction}
\IEEEPARstart{I}{n} order to meet the requirements of real-time monitoring and control, smart grids are highly integrated with advanced information and communication technologies. In recent years, a series of practical applications in smart grids including state estimation have drawn a great deal of attention in research communities due to their vulnerability to cyber-physical attacks. As one of the most threatening attacks, the high-profile false data injection attack (FDIA) aims to mislead the operators into wrong operations by injecting fake measurements, resulting in economic losses and security threats \cite{musleh2019survey}.

As a response, detection methods against FDIA have been widely studied in existing literature, which can be divided into two categories, namely \emph{model-based} and \emph{data-driven} approaches. In \cite{duan2018resilient}, the system's states are estimated via Weighted Least Squares (WLS) estimator and further compared with the received measurements to check the discrepancy revealing potential attacks. A similar detection idea is presented in \cite{manandhar2014detection} with Kalman filter as the estimator in dynamic systems. However, as strictly proved in \cite{liu2011false,liu2017false}, some well-constructed FDIA can bypass the bad data detection (BDD) mechanism so as to achieve stealthy injections. Unlike aforementioned passive model-based algorithms, a proactive detection strategy is proposed in \cite{liu2018reactance,li2020on} by changing the system line parameters via distributed flexible AC transmission systems (D-FACTS) devices to discover attacks constructed with known system line parameters. Also, some data-driven based methods are proposed with the rise of machine learning technologies in the past decades. Among them, supervised learning-based FDIA detection approaches are similar and straightforward from the perspective of idea, i.e., with the measurements as the input, various classifiers including Logistic Regression \cite{esmalifalak2017detecting}, Support Vector Machine \cite{jindal2016decision}, Deep Neural Network \cite{vimalkumar2017big}, Convolutional Neural Network \cite{wang2019detection}, and Recurrent Neural Network \cite{yu2018online} are adopted to distinguish whether the system is suffering from FDIA. In addition, several unsupervised learning-based approaches show their potentialities on FDIA detection. In \cite{zhao2018anomaly}, a deep auto-encoder is utilized while the error between the decoded measurements and the measurements themselves is taken as the detection indicator. The authors of \cite{he2017real} propose to use Deep Belief Network to extract the features of the measurements.       

It is worth noting that most existing FDIA detection studies rely on a key assumption, i.e., the power system can be accurately modeled, or specifically, the system line parameters can be accurately known all the time. Actually, there is still room for discussion on this assumption with two facts in real-world scenarios: 1) As pointed in \cite{Bockarjova2007transmission,Rakpenthai2012state,wang2019dynamic}, system line parameters of smart grids, such as line resistance and reactance, vary with a series of internal/external conditions. 2) When the system is under operation, it is difficult to measure its parameters online. Unfortunately, the above issues are rarely considered in existing publications. In general, the presence of modeling errors in real-world smart grids creates a gap between theoretical analysis and practical implementation.


In order to bridge such a gap, in this paper, we provide a new perspective for FDIA detection considering modeling errors, i.e., \emph{transfer learning}. The key idea is to treat the simulated power system as a \emph{source domain}, which provides abundant simulated attack and normal data to train a deep neural network (DNN) based FDIA detector, while the real-world running system is taken as a \emph{target domain} where sufficient real normal data are collected to fine-tune the DNN. 


The main contributions of this paper is summarized as follows:
      
\begin{enumerate}
	\item[1)] Different from most existing detection methods against FDIA, the modeling errors in transmission lines are considered in this paper. An illustrative example is provided to show that modeling errors can reduce the detection accuracy of conventional FDIA algorithms.
	
	\item[2)] A novel two-stage deep transfer learning-based mechanism is proposed to address the FDIA detection problems considering modeling errors. The performance of the proposed method outperforms the baselines considered in this paper, which reveals the great power of transfer learning on FDIA detection tasks.  
\end{enumerate}

To promote further study of the transfer learning technique for problems involving FDIA, the code used to generate the simulation results in this paper is available online: \url{https://github.com/599143868/DTL-FDIA-TSG}.

The rest of this paper is organized as follows. Section \ref{Section: Problem Formulation} mainly introduces the problem formulation. The proposed deep transfer learning (DTL) based FDIA detection method is illustrated in Section \ref{Section: Transfer Learning Based FDIA Detection Mechanism}. Several case studies are provided in Section \ref{Section: Case Studies} to validate the effectiveness of the proposed approach. At last, some conclusion remarks are given in Section \ref{Section: Conclusion}.  

\section{Problem Formulation}\label{Section: Problem Formulation} 
In this section, we first introduce the AC state estimation in power systems, which is widely employed in most real-world grids. Then, we present a general pattern of stealthy FDIA against AC state estimation that could successfully bypass the bad data detection (BDD) mechanism. Finally, an example is provided to illustrate how modeling errors affect the effectiveness of conventional FDIA detection methods.    
\subsection{AC State Estimation}
The state estimation method is used to supervise the current state of the power system since measuring some specific variables (e.g., voltage phase angle) requires a variety of special sensors, which are usually expensive. Supervisory control and data acquisition system (SCADA) is employed to collect the measured data and transmit them to the control center via communication networks \cite{vimalkumar2017big}.

In an AC state estimation model, the relationship between measurements $\bm{z} \in \mathbb{R}^m$ and estimated states $\bm{x} \in \mathbb{R}^n$ can be represented as follows:
\begin{align}\label{Eq: AC state estimation}
	\bm{z} = h(\bm{x})+\bm{e}
\end{align}     
where $h(\cdot)$ denotes the nonlinear dependency between $\bm{z}$ and $\bm{x}$, which is determined by the system line parameters such as system topology and line admittance, $\bm{e} \in \mathbb{R}^m$ is the measurement error. In this paper, the measurements $\bm{z}$ contain bus active/reactive power injection $P_i$, $Q_i$, and line flow $p_{ij}$, $q_{ij}$, while the states $\bm{x}$ consist of bus voltage amplitude $V_i$ and phase angle $\theta_i$. Thus, (\ref{Eq: AC state estimation}) could be developed from the following equations \cite{liu2017false}:
\begin{align}\label{Eq: line flow}
	&p_{ij} = V_i^2 g_{ij} - V_iV_j[g_{ij}\cos(\theta_i-\theta_j)+b_{ij}\sin(\theta_i-\theta_j)] \nonumber \\
	&q_{ij} = -V_i^2 b_{ij} - V_iV_j[g_{ij}\sin(\theta_i-\theta_j)-b_{ij}\cos(\theta_i-\theta_j)] \nonumber \\
	&P_i = \sum_{j\in \mathcal{N}_i}p_{ij},~Q_i = \sum_{j\in \mathcal{N}_i}q_{ij},~
\end{align}
where $\mathcal{N}_i$ denotes the bus neighbors of bus $i$. When the control center receives the measured data $\bm{z}$ from SCADA, it estimates $\bm{x}$ as $\hat{\bm{x}}$ by solving equation (\ref{Eq: AC state estimation}).

\subsection{A General Pattern of FDIA}
The purpose of FDIA is to mislead the operator to estimate a compromised $\hat{\bm{x}}_a=\hat{\bm{x}}+\bm{c}$ by changing $\bm{z}$ to a constructed false injection $\bm{z}_a=\bm{z}+\bm{a}$, where $\bm{a}\in \mathbb{R}^m$ is an injected attack vector \cite{li2019online}.

Some defenders use BDD mechanism to detect potential attacks, which computes the Euclidean norm of the residual $\bm{r}=\bm{z}-h(\hat{\bm{x}})$ and compares it with a preset threshold $\tau$. If $\left\|\bm{r}\right\|_2\le \tau$, the measurements $\bm{z}$ are considered trustworthy; otherwise, they are considered attacked.  

Different from conventional injection attacks, stealthy FDIA constructs $\bm{a}$ to bypass the BDD mechanism via the following strategy \cite{liu2011false}:
\begin{align}\label{Eq: construct a}
	\bm{a} = h(\bm{x}+\bm{c})-h(\bm{x})	
\end{align}
In this way, we readily find that the Euclidean norm of the residual is unchanged:
\begin{align}\label{Eq: FDIA}
	\left\|\bm{r}_a\right\|_2 &=	\left\|\bm{z}_a-h(\hat{\bm{x}}_a)\right\|_2	= \left\|\bm{z}+\bm{a}-h(\hat{\bm{x}}+\bm{c})\right\|_2  \nonumber \\
	&=\left\|\bm{z}-h(\hat{\bm{x}}) \right\|_2 = \left\|\bm{r} \right\|_2
\end{align} 
and thus the constructed stealthy FDIA can circumvent BDD. 

\textcolor[RGB]{0,32,128}{\textmd{\textsf{\emph{Remark 1:}}}} We should mention that there have been some studies proposing a series of FDIA based on different attack objectives and also require the attacker for different levels of system knowledge \cite{liu2017false,wang2019dynamic,yan2020false}. For the sake of simplicity, a general pattern of stealthy FDIA described in (\ref{Eq: construct a})-(\ref{Eq: FDIA}) is considered in this paper.

\subsection{Limitations of Existing Detection Methods Considering Model Errors}
Most existing FDIA detection methods rely on a key assumption, i.e., the real-world power system line parameters (e.g., line admittance) are known precisely, which introduces two practical issues into our sight. First, \emph{are power system line parameters time-varying?} The answer should be positive. As pointed out in \cite{Bockarjova2007transmission,Rakpenthai2012state,wang2019dynamic}, the system line parameters vary with weather conditions, environmental conditions, and equipment age. Second, \emph{are power system line parameters real-time available?} The answer should be negative since precise measurement of system line parameters is not practical when the system is running. Therefore, recalling FDIA detection methods relying on precisely known system line parameters, we readily find that they are difficult to achieve in practical scenarios. Then, it is natural to ask a question: \emph{What is the effect of these unknown changes on normal FDIA detection algorithms? An illustrative example is given below to answer this question.}

\begin{figure}[!t]
	\centering
	\includegraphics[width=2.5in]{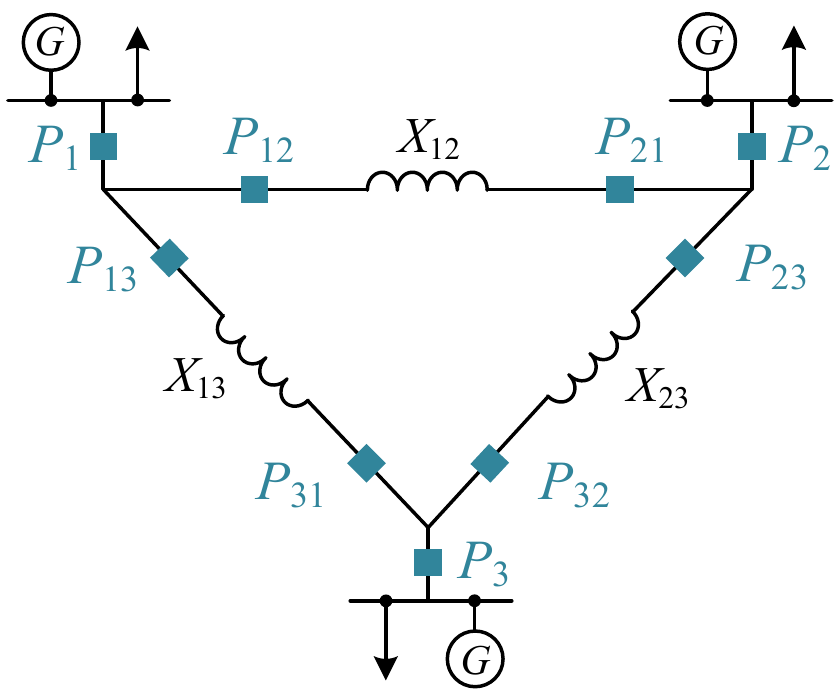}
	\caption{Illustrative 3-bus system, where the blue squares denote meters.}
	\label{Fig: sample}
	\vspace{-1.5em}
\end{figure}

\emph{Illustrative Example:} As shown in Fig. \ref{Fig: sample}, the considered power system includes three buses and three branches. The resistances on the three branches are assumed to be 0, while the reactances are set to $X_{12}$=0.0281 $\Omega$, $X_{13}$=0.0304 $\Omega$, $X_{23}$=0.0108 $\Omega$. 
	
For the sake of simplicity and clarity, in this example, the AC model (\ref{Eq: AC state estimation}) is reduced to the following linear regression form
\begin{align}\label{Eq: linear regression}
\bm{z} = \bm{Hx} + \bm{e}
\end{align}
with $\bm{H} \in \mathbb{R}^{m\times n}$ derived from the following equation
\begin{align}\label{Eq: line flow2}
	P_{ij} = \frac{1}{X_{ij}}(\theta_i-\theta_j)
\end{align}

The state variables are chosen as $\bm{x}=[\theta_1,\theta_2,\theta_3]^T$ and meter measurements as $\bm{z}=[P_{12},P_{13},P_{23}]^T$. Then according to (\ref{Eq: linear regression}) and (\ref{Eq: line flow2}), $\bm{H}$ can be expressed as   
\begin{align}\label{Eq: H}
\bm{H}=	\tiny\begin{bmatrix}
	\frac{1}{X_{12}} & -\frac{1}{X_{12}} & 0 \\
	\frac{1}{X_{13}} & 0 & -\frac{1}{X_{13}}  \\
	0 & \frac{1}{X_{23}} & -\frac{1}{X_{23}}   	
\end{bmatrix}=\tiny
\begin{bmatrix}
	35.59 & -35.59 & 0 \\
	32.89 & 0 & -32.90\\
	0 & 92.59 & -92.59   	
\end{bmatrix}
\end{align}

Assuming the state variables $\bm{x}=[0,-0.0106,-0.0006]^T$ and the measurement error $\bm{e}\sim (0,0.0001^2)$, then the defender can meter a normal measurement
\begin{align}\label{Eq: calculate z}
\bm{z} = \tiny
\begin{bmatrix}
	35.59 & -35.59 & 0 \\
	32.89 & 0 & -32.90\\
	0 & 92.59 & -92.59   	
\end{bmatrix}\tiny\begin{bmatrix}
0 \\ -0.0106 \\ -0.0006 
\end{bmatrix}+\tiny\begin{bmatrix}
-0.00010 \\ -0.00011 \\~~~0.00013 
\end{bmatrix}=\tiny\begin{bmatrix}
~~~0.3787 \\ ~~~0.0221 \\-0.9231 
\end{bmatrix}.	
\end{align}
Based on (\ref{Eq: linear regression}) and WLS estimator, the state variables are estimated as follows
\begin{align}\label{Eq: estimate x}
\bm{\hat{x}}=(\bm{H}^T\bm{R}\bm{H})^\dagger \bm{H}^T\bm{R}\bm{z}=\tiny\begin{bmatrix}
	0 \\ -0.0106 \\ -0.0007 
\end{bmatrix}
\end{align}
where $\bm{R}=0.0001\times \bm{I}$ is the covariance matrix of $\bm{e}$. Next, with the obtained (\ref{Eq: H}), (\ref{Eq: calculate z}), (\ref{Eq: estimate x}), and according to the criterion of BDD, the residual is computed as
\begin{align}\label{Eq: compute r}
\left\|\bm{r}\right\|_2=\left\|\bm{z}-\bm{H}\hat{\bm{x}}\right\|_2=	4.68e-5.
\end{align}
Let the preset threshold $\tau$ be 10 times the standard deviation of $\bm{e}$, i.e., $\tau=0.001$, we easily find that the residual is much smaller than the threshold, i.e., $\left\|\bm{r}\right\|_2 < \tau$. In this way, the normal measurement $\bm{z}$ is correctly detected as normal.

However, what would happen when there are modeling errors in transmission line parameters?

Here, we consider modeling errors within 10\%. Specifically, the aforementioned $X_{12},X_{13},X_{23}$ are treated as nominal transmission line parameters which are used in the simulation, whereas those in real world are established by adding perturbations within 10\%: $X_{12}^*$=0.0260 $\Omega$, $X_{13}^*$=0.0287 $\Omega$, $X_{23}^*$=0.0116 $\Omega$. Therefore, the Jacobian matrix in real world is expressed as   
\begin{align}\label{Eq: compute H^star}
\bm{H^*}=	\tiny\begin{bmatrix}
		\frac{1}{X_{12}^*} & -\frac{1}{X_{12}^*} & 0 \\
		\frac{1}{X_{13}^*} & 0 & -\frac{1}{X_{13^*}}  \\
		0 & \frac{1}{X_{23}^*} & -\frac{1}{X_{23}^*}   	
	\end{bmatrix}=\tiny
	\begin{bmatrix}
		38.49 & -38.49 & 0 \\
		34.88 & 0 & -34.88\\
		0 & 86.20 & -86.20   	
\end{bmatrix}	
\end{align}
Then the measurements metered by the defender are as follows
\begin{align}\label{Eq: compute z^star}
	\bm{z^*} = \tiny
		\begin{bmatrix}
		38.49 & -38.49 & 0 \\
		34.88 & 0 & -34.88\\
		0 & 86.20 & -86.20   	
		\end{bmatrix}\tiny\begin{bmatrix}
			0 \\ -0.0106 \\ -0.0006 
		\end{bmatrix}+\tiny\begin{bmatrix}
			-0.00010 \\ -0.00011 \\~~~0.00013 
		\end{bmatrix}=\tiny\begin{bmatrix}
			~~~0.4096 \\ ~~~0.0234 \\-0.8594 
		\end{bmatrix},	
\end{align}
and the state variables are estimated as follows
\begin{align}\label{Eq: estimate x^star}
	\bm{\hat{x}^*}=(\bm{H}^T\bm{R}\bm{H})^\dagger \bm{H}^T\bm{R}\bm{z^*}=\tiny\begin{bmatrix}
			~~~0.3863 \\ ~~~0.0486 \\ -0.8683 
	\end{bmatrix}.
\end{align}
Further, the residual is computed as
\begin{align}\label{Eq: compute r^star}
	\left\|\bm{r}^*\right\|_2=\left\|\bm{z}^*-\bm{H}\hat{\bm{x}}^*\right\|_2=	0.0355.
\end{align}

It can be observed that due to the presence of modeling errors in transmission line parameters, the obtained residual is much larger than the preset threshold, i.e., $\left\|\bm{r}^*\right\|_2 > \tau$, entailing that the normal measurement $\bm{z}$ is detected as an attack by BDD. The rationale behind this result is that the $\bm{H}$ used to estimate the state variables in (\ref{Eq: estimate x^star}) is inconsistent with the $\bm{H}^*$ providing the measurements in (\ref{Eq: compute z^star}), which is essentially due to the fact that the transmission line parameters in power systems are time-varying and difficult to measure in real time. In fact, similar results are observed in other model-based and data-driven methods, which are also shown in our case studies. Now, from the perspective of defenders, we are facing a challenging question: \emph{How can we achieve high-precision FDIA detection when the system line parameters cannot be accurately known?} Such a question motivates us to propose our transfer learning-based approach. 
   
\section{Transfer Learning Based FDIA Detection Mechanism}\label{Section: Transfer Learning Based FDIA Detection Mechanism}
In this section, we provide a new perspective, i.e., \emph{transfer learning}, to achieve high-precision detection of stealthy FDIA attacks with system line parameters that cannot be accurately known. As an overview, in our proposed detection mechanism, we aim to adapt a deep neural network trained on the simulation data for the real-world data. Considering the performance of a transfer approach depends on a series of operational details, we will describe the proposed detection mechanism in terms of data collection, transfer design, and network design. 
\begin{figure*}[!t]
	\centering
	\includegraphics[width=7.0in]{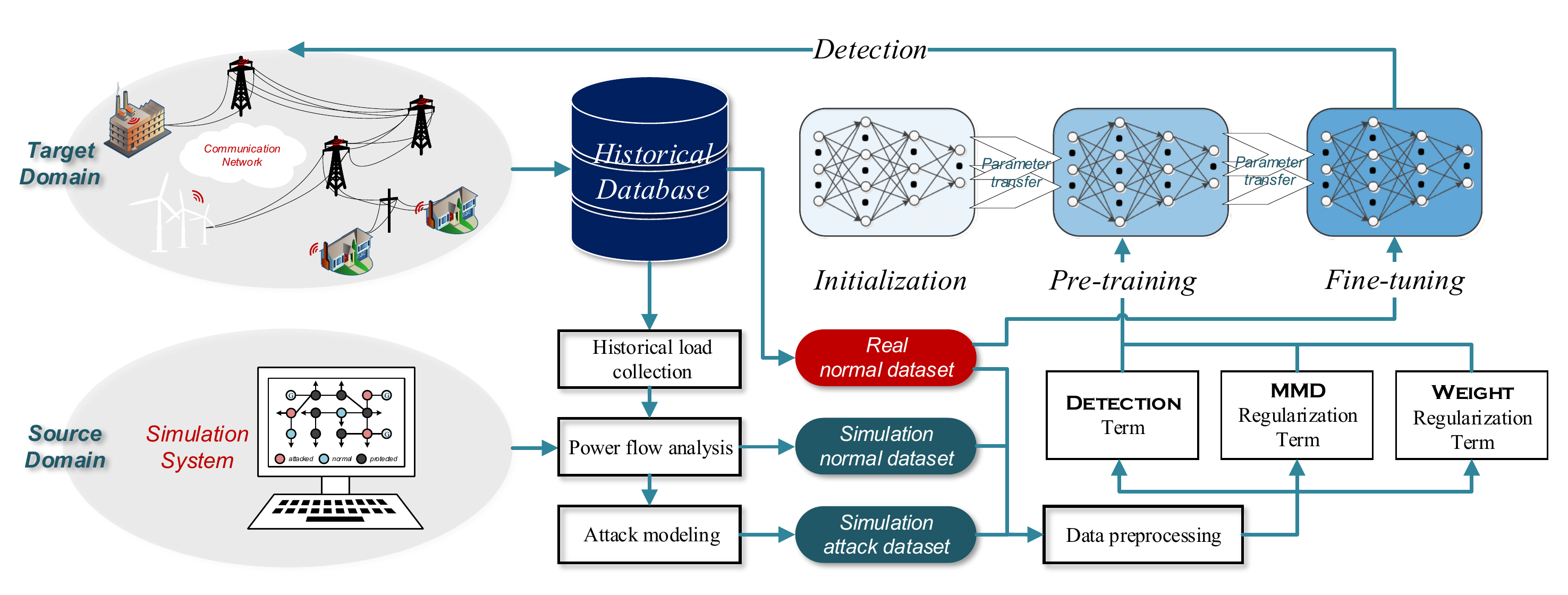}
	\caption{Proposed transfer learning based FDIA detection mechanism.}
	\label{Fig: Transfer design}
	\vspace{-1.5em}
\end{figure*}

\subsection{Data Collection}
First of all, we should be clear about what data we have as defenders. First, we denote the system line parameters as $H$. We can employ this ideal $H$ in a simulation environment to generate sufficient \emph{simulated} normal/attack measurement data. Second, even if the system line parameters at the current moment, denoted as $H^*$, cannot be accurately known, we can collect the \emph{real} normal measurement data online through the deployed sensors and communication network.  

From the perspective of transfer learning, recalling the fact that $H$ is precisely known while $H^*$ is not, it is natural to treat $H$ as the \emph{source domain} and $H^*$ as the \emph{target domain}. For the simulation (source domain) dataset, different bus load demand vectors $\bm{P}_{S_d}$ and $\bm{Q}_{S_d}$ are required to simulate the practical load conditions. Based on this, we can obtain the simulated normal measurement dataset $\mathcal{D}_{S_N}=\{\bm{P}_{S_d}^{(k)},\bm{Q}_{S_d}^{(k)},\bm{P}_{S_N}^{(k)},\bm{Q}_{S_N}^{(k)},\bm{p}_{S_N}^{(k)},\bm{q}_{S_N}^{(k)}\}_{k=1}^{K_{S_N}}$ via existing power flow algorithms, and generate the corresponding attack dataset $\mathcal{D}_{S_A}=\{\bm{P}_{S_d}^{(k)},\bm{Q}_{S_d}^{(k)},\bm{P}_{S_A}^{(k)},\bm{Q}_{S_A}^{(k)},\bm{p}_{S_A}^{(k)},\bm{q}_{S_A}^{(k)}\}_{k=1}^{K_{S_A}}$ by setting a series of attack vectors (different attack positions and intensities) via (\ref{Eq: construct a})-(\ref{Eq: FDIA}). Note that $k$ indexes the data sample, $K_{S_N}$ and $K_{S_A}$ denote the number of normal samples and attack samples, respectively. For better training performance, the number of generated normal samples and attack samples in the final constituted simulation dataset $\mathcal{D}_{S}=\{\mathcal{D}_{S_N},\mathcal{D}_{S_A}\}$ should be balanced, that is, $K_{S_N}=K_{S_A}$. For the real (target domain) dataset, representative bus load demand vectors $\bm{P}_{T_d}$ and $\bm{Q}_{T_d}$ are required for better transfer performance. In this way, the real normal measurement dataset $\mathcal{D}_{T_N}=\{\bm{P}_{T_d}^{(k)},\bm{Q}_{T_d}^{(k)},\bm{P}_{T_N}^{(k)},\bm{Q}_{T_N}^{(k)},\bm{p}_{T_N}^{(k)},\bm{q}_{T_N}^{(k)}\}_{k=1}^{K_{T_N}}$ can be collected. Since real attack data are difficult to collect, the final constituted real dataset is $\mathcal{D}_T=\mathcal{D}_{T_N}$. Note that $K_{S_N}$, $K_{S_A}$, and $K_{T_N}$ can all be huge (e.g., millions) due to the following two reasons: 1) The cost of collecting these data is very cheap. 2) Training tasks on large-scale datasets no longer require too expensive hardware support due to the development of training acceleration techniques (e.g., mini-batch methodology, batch normalization) in recent years.  
  
\subsection{Transfer Design}
Due to the fact that our real (target domain) data do not include any real attack samples but only real normal samples, the transfer method should be designed with great care to fully utilize the information we have. In general, our designed transfer method is divided into two consecutive training stages: the pre-training stage and the fine-tuning stage. Note that these two stages have their unique optimization objectives and will be introduced in sequence.

\subsubsection{Stage 1: Pre-training} Some existing transfer methods are straightforward, e.g., only source domain data are required in the pre-training stage as each category in their target domain dataset has sufficient data and labels for supervised fine-tuning. Considering our target domain lacks available attack samples, both empirically and experimentally, it is difficult to achieve satisfactory transfer performance with only source domain data used in this pre-training stage. To address this issue, we incorporate the idea of \emph{domain adaptation} into conventional supervised learning and propose the training strategy in this pre-training stage, which implies the following three main optimization objectives: 
\begin{enumerate}
	\item[1)] Training a DNN to classify the simulation (source domain) dataset as correctly as possible.  
	
	\item[2)] The trained DNN needs to provide considerable adaptability on the real (target domain) dataset as an initial network in the fine-tuning stage. 
	
	\item[3)] The overfitting effect should be mitigated as much as possible. 
\end{enumerate}

For the first optimization objective, a conventional cross-entropy loss function is employed to drive the DNN-based classifier to approach the relationship between the source domain data $\mathcal{D}_S$ and their corresponding labels:
\begin{align}\label{Eq: loss_cl}
	\mathcal{L}_{cl} = -\frac{1}{K_S}\sum_{k=1}^{K_S}(y_a\log \tilde{y}_{a}(\mathcal{D}_S^{(k)},\Theta)+y_n\log \tilde{y}_{n}(\mathcal{D}_S^{(k)},\Theta))	
\end{align}
where $K_S$ denotes the number of source domain samples contained in a single training batch, $\tilde{y}_{a}(\mathcal{D}_S^{(k)},\Theta)$ and $\tilde{y}_{n}(\mathcal{D}_S^{(k)},\Theta)$ are two outputs of the DNN classifier $\Theta$, which represent the predicted possibility that the current sample $\mathcal{D}_S^{(k)}$ is attacked/normal. If $\tilde{y}_{a}(\mathcal{D}_S^{(k)},\Theta)>\tilde{y}_{n}(\mathcal{D}_S^{(k)},\Theta)$, $\mathcal{D}_S^{(k)}$ is predicted as an attack sample; otherwise, it is predicted to be normal. $y_a$ and $y_n$ indicate the corresponding labels. In this work, $(y_a,y_n)=(1,0)$ for an attack sample, $(y_a,y_n)=(0,1)$ for a normal one. Note that (\ref{Eq: loss_cl}) is a typical supervised term since the labels for normal samples and attack samples in $\mathcal{D}_S$ are required. 

For the second optimization objective, motivated by the idea of domain adaptation, the \emph{Maximum Mean Discrepancy} (MMD) term is utilized to measure the distribution discrepancy between the sample means of $\mathcal{D}_S$ and $\mathcal{D}_T$ in a reproducing kernel Hilbert space $\mathcal{H}$:

\begin{align}\label{Eq: loss_MMD}
	\mathcal{L}_{MMD} = \left \| \frac{1}{K_S}\sum_{k=1}^{K_S} \tilde{y}_{\mathcal{J}}(\mathcal{D}_S^{(k)},\Theta_{\mathcal{J}})-\frac{1}{K_T}\sum_{k=1}^{K_T} \tilde{y}_{\mathcal{J}}(\mathcal{D}_T^{(k)},\Theta_{\mathcal{J}}) \right \|_\mathcal{H} 
\end{align}
where $K_T$ denotes the number of target domain samples contained in a single training batch. $\Theta_{\mathcal{J}}$ represents the first $J$ layers of DNN with their last layer as the output. $\tilde{y}_{\mathcal{J}}(\cdot,\Theta_{\mathcal{J}})$ indicates the feature of the input expressed by $\Theta_{\mathcal{J}}$. As pointed out in \cite{gretton2006kernel}, by minimizing (\ref{Eq: loss_MMD}), the features of $\mathcal{D}_S$ and $\mathcal{D}_T$ can be drawn closer in $\mathcal{H}$. Note that (\ref{Eq: loss_MMD}) is a kind of unsupervised term as no labels are required.   

For the third optimization objective, a weights regularization term is adopted for mitigating the overfitting effect. As pointed out in \cite{lu2016deep}, the elements of $\Theta_\mathcal{J}$ possibly approach zero so as to make the expressed features meet the minimization of (\ref{Eq: loss_MMD}). Therefore, the regularization term is expressed as follows:
\begin{align}\label{Eq: loss_Theta}
	\mathcal{L}_{\Theta} = \exp(-\left\| \Theta_\mathcal{J} \right\|_F)
\end{align}
where $\left\| \cdot \right\|_F$ denotes \emph{Frobenius norm}.

By integrating (\ref{Eq: loss_cl}), (\ref{Eq: loss_MMD}), and (\ref{Eq: loss_Theta}), the final objective function in this stage is represented as follows:
\begin{align}\label{Eq: loss}
	\mathcal{L} = \mathcal{L}_{cl}+\lambda \mathcal{L}_{MMD}+\mu \mathcal{L}_{\Theta} 
\end{align}
where $\lambda$ and $\mu$ are related positive trade-off hyper-parameters, which are usually selected empirically and through the ``trial and error" process. In order to further avoid overfitting of DNN, the whole source domain dataset $\mathcal{D}_S$ is divided into a training set and a validation set at a ratio of 7:3 with reference to commonly use training strategies \cite{lecun2015deep}. To further improve the training performance, the \emph{Xavier} normal initializer is adopted for the initialization of the network weights $\Theta$ \cite{glorot2010understanding}. The training process in this stage stops when both the classification accuracy of DNN on the training set and that on the validation set exceed a preset threshold. 

\subsubsection{Stage 2: Fine-tuning} Once the pre-training stage is completed, we use the real (target domain) dataset $\mathcal{D}_T$ and their corresponding labels to fine-tune the DNN weights $\Theta$ supervisedly. A conventional cross-entropy loss function is adopted as the objective function in this stage.
\begin{align}\label{Eq: loss_}
	\mathcal{L}_{cl} = -\frac{1}{K_T}\sum_{k=1}^{K_T}(y_a\log \tilde{y}_{a}(\mathcal{D}_T^{(k)},\Theta)+y_n\log \tilde{y}_{n}(\mathcal{D}_T^{(k)},\Theta))	
\end{align}

Note that the learning rate should be set very small to avoid excessive loss of the pre-trained model. Similar to the operations on $\mathcal{D}_S$, $\mathcal{D}_T$ should be also divided into a training set and a validation set. The training process stops when the number of training iterations exceeds a preset threshold.

The whole designed transfer method is illustrated in Fig. \ref{Fig: Transfer design}. Note that the training set originated from $\bm{D}_S$ has been normalized into $[-1,1]$ to accelerate the training process, and the corresponding normalization is also applied to other involved training/validation set. The optimizer we employ for the two training tasks is \emph{adam}, which has been widely acclaimed by researchers recently due to its outstanding efficiency and adaptability for a variety of complex optimization problems \cite{kingma2014adam}. Besides, the batch size and learning rate are selected by cross validation.  

\textcolor[RGB]{0,32,128}{\textmd{\textsf{\emph{Remark 2:}}}} Similar to most existing learning-based FDIA detection approaches, we assume that the power grid's physical \emph{topology} does not change dramatically within a short time. Once the topology is updated, it is suggested to re-transfer with real data collected from the new system. Note that the time cost of such a transfer operation is fairly cheap, which will be verified in our simulation results.
  
\subsection{Network Design}
\begin{figure}[!t]
	\centering
	\includegraphics[width=3.4in]{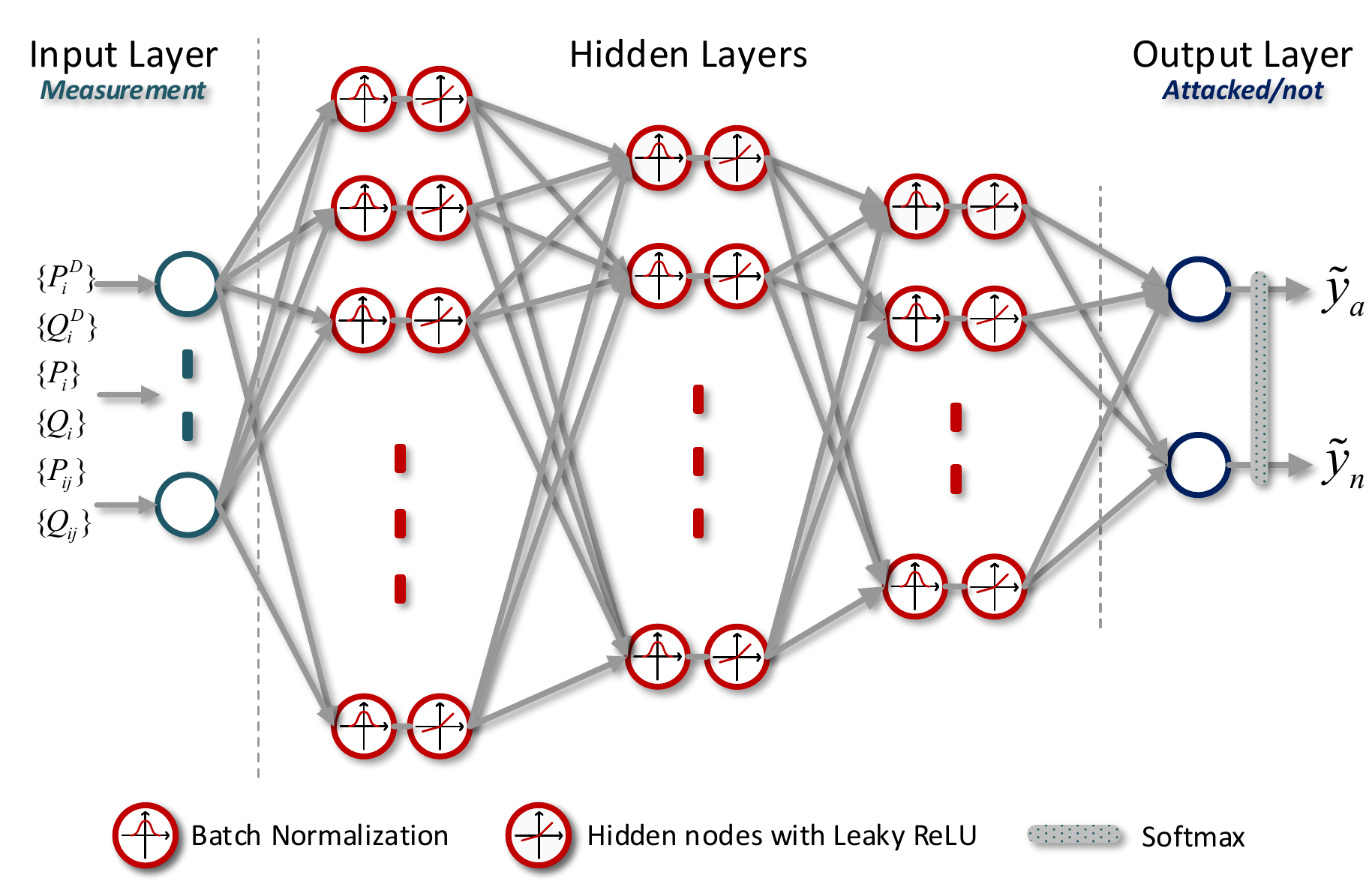}
	\caption{Proposed transfer learning based FDIA detection mechanism: Network structure design.}
	\vspace{-1.5em}
	\label{Fig: Network design}
\end{figure}

As shown in Fig. \ref{Fig: Network design}, we adopt a fully-connected DNN structure, which contains one input layer, multiple hidden layers, and one output layer. In the output layer, to capture the false components that may exist in the measurements, a preliminary strategy is to directly treat $\bm{z}$ as the input of the classifier, which is also widely adopted in most existing learning-based FDIA detection approaches \cite{he2017real,yu2018online,aboelwafa2020machine,wang2020locational}. However, let us consider an important question: \emph{Can we guarantee that the attack sample $\bm{z}_A$ under a certain load condition will not be the same as the normal sample $\bm{z}_N$ under another load condition?} To our best knowledge, the answer should be negative. In other words, if we only treat $\bm{z}$ as the input of the network, there is indeed the possibility of collecting two identical samples with different labels. Such a behavior undoubtedly confuses the training optimizer so that the midpoint of the different labels is regarded as the real working label, which considerably damages the performance of the classifier. Therefore, the bus load demand vectors $\bm{P}_d$ and $\bm{Q}_d$, in addition to the conventional measurements $\bm{P}$, $\bm{Q}$, $\bm{p}$, $\bm{q}$, are taken as the inputs of our designed network. In the hidden layers, we select the Leaky Rectified Linear Unit (Leaky ReLU) as the nonlinear activation function for each hidden node. What's more, we add a batch normalization (BN) layer before each hidden layer to accelerate the training process and mitigate the overfitting effect. In the output layer, a softmax function is utilized to predict the current sample is a normal/attack one.


\textcolor[RGB]{0,32,128}{\textmd{\textsf{\emph{Remark 3:}}}} It is the modeling errors in power system transmissions lines that motivate us to distinguish the running system's line parameters $H^*$ from its simulation system's line parameters $H$, and develop our two-stage DTL-based approach which aims at adapting DNN trained with the simulation system to the running system so as to track the varying modeling errors online. In contrast, most conventional FDIA detection approaches neglect the presence of modeling errors in power system transmissions lines and default $H^*=H$. As demonstrated in the Case Studies presented in the following section,  such a two-stage DTL-based method outperforms the conventional deep learning-based approaches. Especially, more improvements are evident with increasing modeling errors. This can be seen as a significant contribution of this paper. 

\textcolor[RGB]{0,32,128}{\textmd{\textsf{\emph{Remark 4:}}}} As the mapping relationship to be expressed by the classifier is fairly complicated, it is necessary to introduce the BN layer into our network. A lot of literature has studied how BN accelerates the training of DNN and avoids vanishing gradient problem \cite{ioffe2015batch,santurkar2018does,bjorck2018understanding}. What we want to emphasize here is that as an approach that may be applied to the practical power grid, reducing training time and avoiding the risk of overfitting will bring substantial convenience to the operators. 

\section{Case Studies}\label{Section: Case Studies}
In this section, we assess the performance of our proposed deep transfer learning based FDIA detection mechanism on the IEEE 14-bus and 118-bus systems. We first introduce how we prepare the representative simulated/real (i.e., source/target domain) datasets in Section \ref{Section: Data Generation}. Then we show the convergence performance of the proposed two-stage DTL-based algorithm with different power system and modeling error levels in Section \ref{Section: Network Transfer}. Finally, the detection accuracy and missing alarm rate of the proposed method are investigated and compared with several baselines in Section \ref{Section: Detection Results}.

Our simulations are carried out on a PC equipped with an 8-core Intel i7-9700 3.0GHz CPU and two 8GB RAMs. The datasets are generated in MATLAB R2018a while the involved transfer tasks are implemented in Python 3.7.2 and Tensorflow 1.13.1. 

All the MATLAB and Python codes (including the proposed method and all involved baselines) are available online: \url{https://github.com/599143868/DTL-FDIA-TSG}.

\subsection{Data Generation}\label{Section: Data Generation}
Here we take the IEEE 14-bus Case as an instance (the IEEE 118-bus Case is also established in the same way). Firstly, the standard IEEE-14 bus system is treated as the simulation system (source domain) $H$ in our work. Then the real-world power system (target domain) $H^*$ is built by revising the resistance and reactance in the lines to new values within a \emph{percentage} $\delta$. The impact of different $\delta$ on the effectiveness of the proposed method will be illustrated and analyzed in our results. For the involved bus load demand vectors $\{\bm{P}_d,\bm{Q}_d\}$, we randomly generate 100 base load demand conditions.  

\subsubsection{Source domain dataset}
Then for each base load demand condition, 5000 specific load demand conditions are randomly generated, ranging from 50\% to 150\% of their base condition. In this way, 500000 bus load demand vectors are prepared. For each bus load demand vector $\{\bm{P}_{S_d}^{(k)},\bm{Q}_{S_d}^{(k)}\}$, a normal measurement $\{\bm{P}_{S_N}^{(k)},\bm{Q}_{S_N}^{(k)},\bm{p}_{S_N}^{(k)},\bm{q}_{S_N}^{(k)}\}$ is obtained by running MATPOWER with $H$ as the system line parameters. The measurement noise is randomly selected within a \emph{percentage} $\sigma$=1\% of the measurement \cite{yu2018online}. Also, an attack measurement $\{\bm{P}_{S_A}^{(k)},\bm{Q}_{S_A}^{(k)},\bm{p}_{S_A}^{(k)},\bm{q}_{S_A}^{(k)}\}$ is synchronously constructed, where the attacked bus is randomly selected in \{bus 2, bus 3, bus 9\} and the corresponding attack intensity is randomly selected from \{10\%, 20\%, 30\%\}. The rationale behind such settings mainly includes the following two points: 1) The number of tampered variables should be limited due to the presence of attack cost \cite{margossian2019partial}. 2) 30\% is taken as the upper bound of the attack intensity since the normal distribution of the measurements is well known by the operators/defenders in the control center, indicating an attack vector with an overly large attack intensity is fairly doubtful. In this way, the collection of the simulated (source domain) dataset $\bm{D}_S$ is completed, which has 1e6 simulated samples containing normal samples and attack samples with a ratio of 1:1.

\subsubsection{Target domain dataset}      
In a similar way, a real (target domain) dataset $\bm{D}_T$ containing only 1e6 normal samples is collected with $H^*$ as the system line parameters. Care has to be exercised here as an additional test dataset $\bm{D}_T^*$ is generated to assess the transfer performance of our proposed detection mechanism, i.e., it does not participate in any training process but only provides test results. The generation of $\bm{D}_T^*$ also uses $H^*$ as the system line parameters, and the other details are similar to those of $\bm{D}_S$, involving the same dataset size and sample proportion.
\begin{figure}[!t]
	\centering
	\subfigure[Pre-training ($\delta$=10\%).\label{Fig: 14bus_10_stage1}]{\includegraphics[width=1.7in]{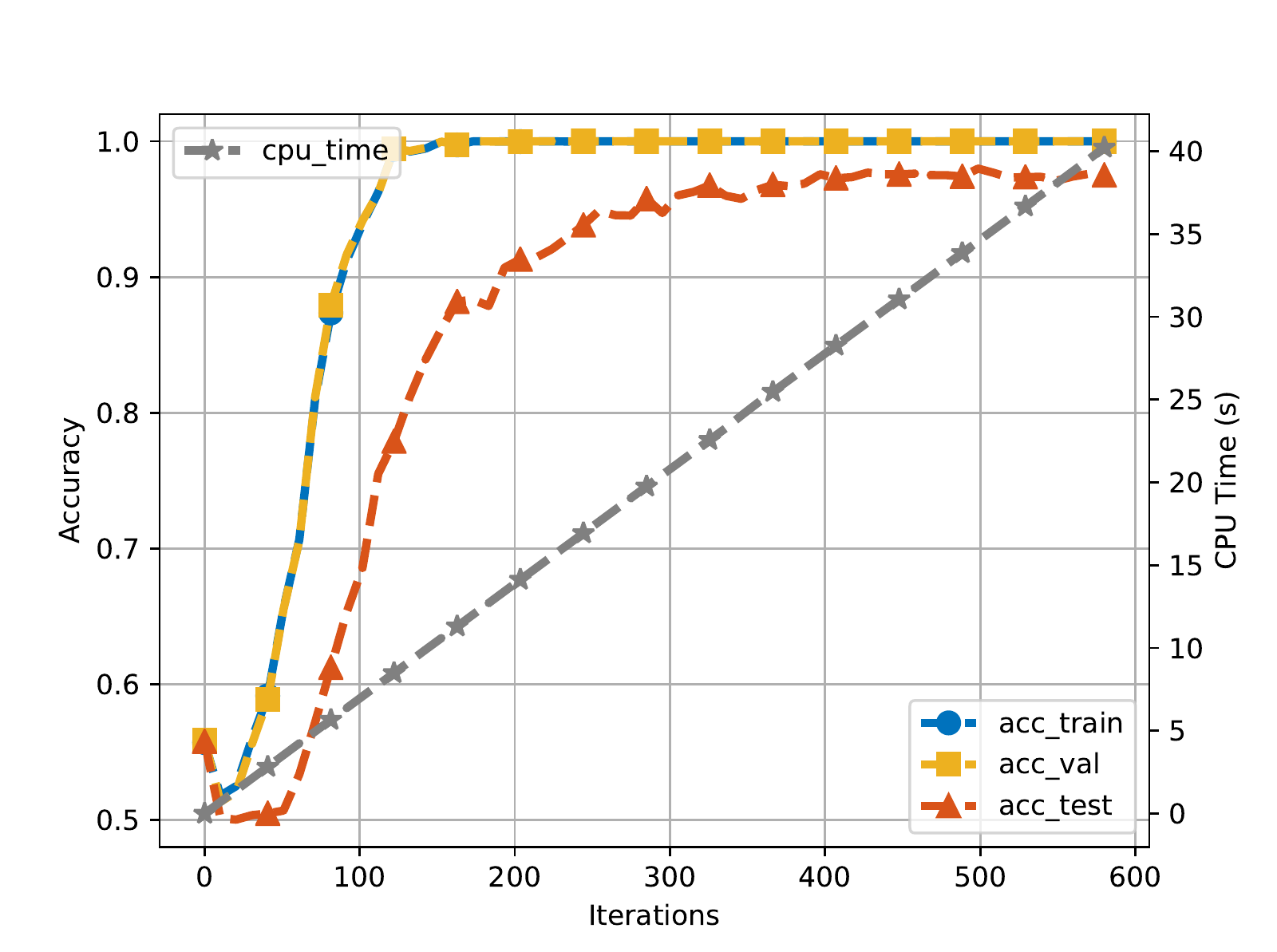}}
	\subfigure[Fine-tuning ($\delta$=10\%).\label{Fig: 14bus_10_stage2}]{\includegraphics[width=1.7in]{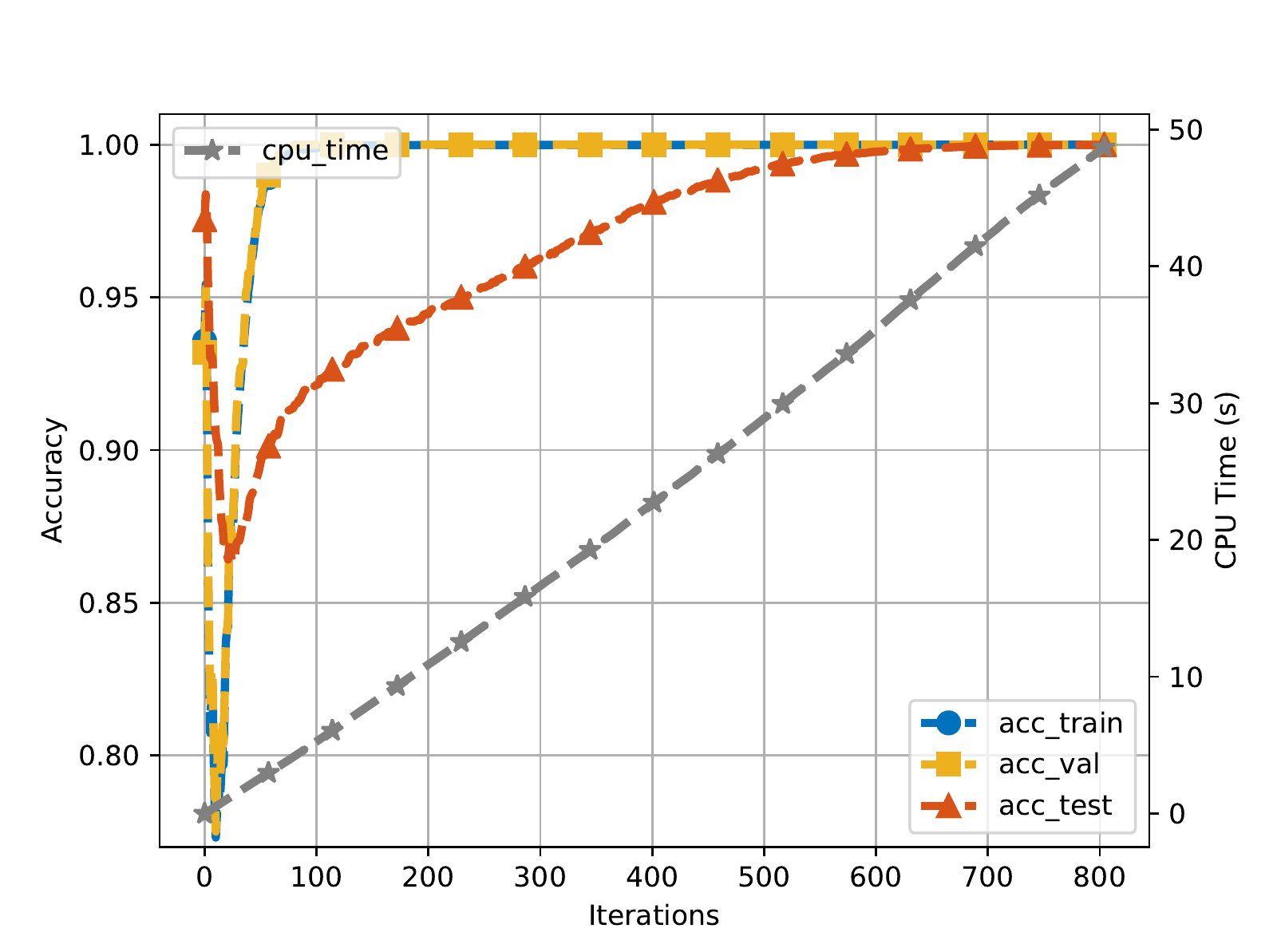}}
	\subfigure[Pre-training ($\delta$=30\%).\label{Fig: 14bus_30_stage1}]{\includegraphics[width=1.7in]{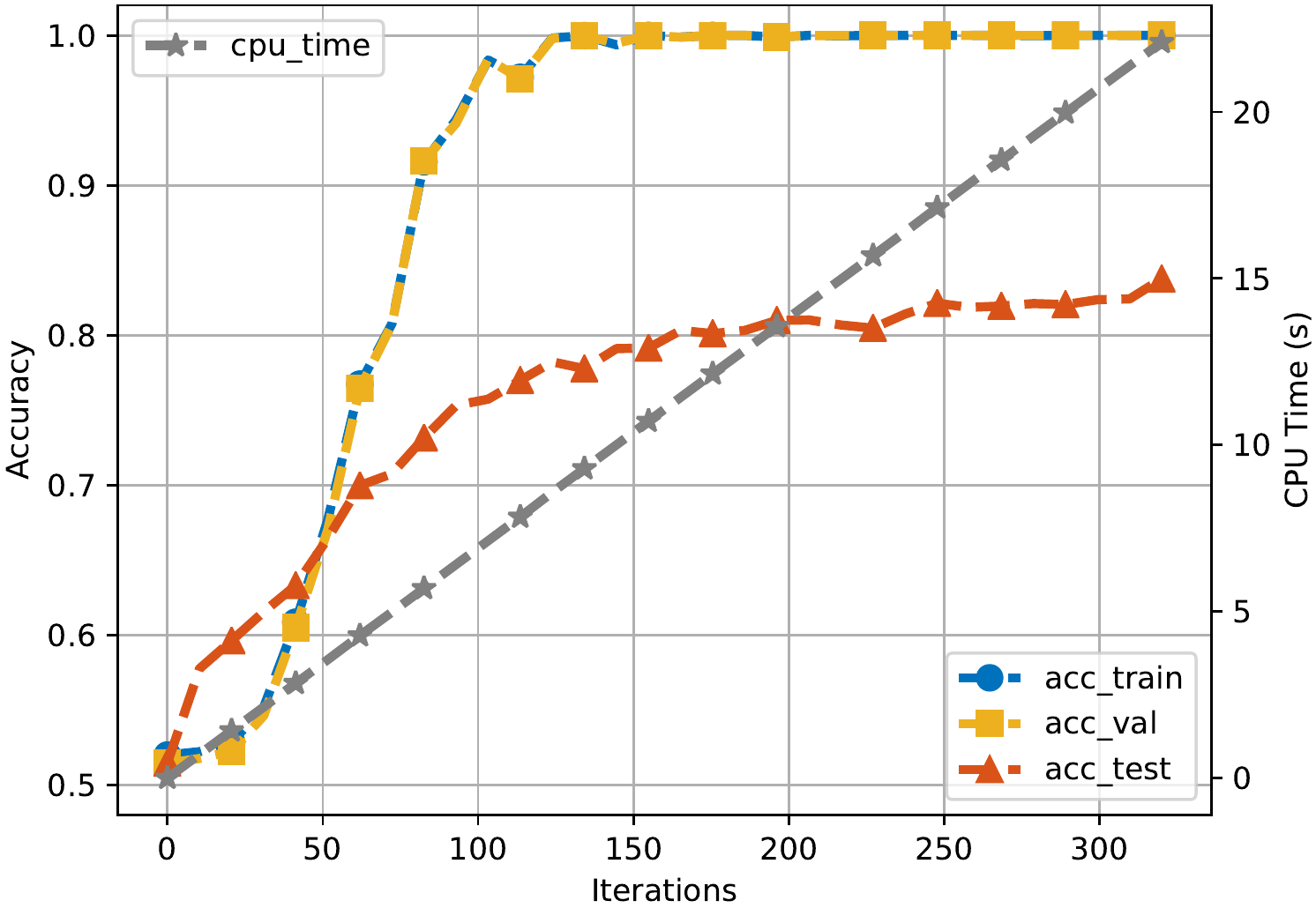}}
	\subfigure[Fine-tuning ($\delta$=30\%).\label{Fig: 14bus_30_stage2}]{\includegraphics[width=1.7in]{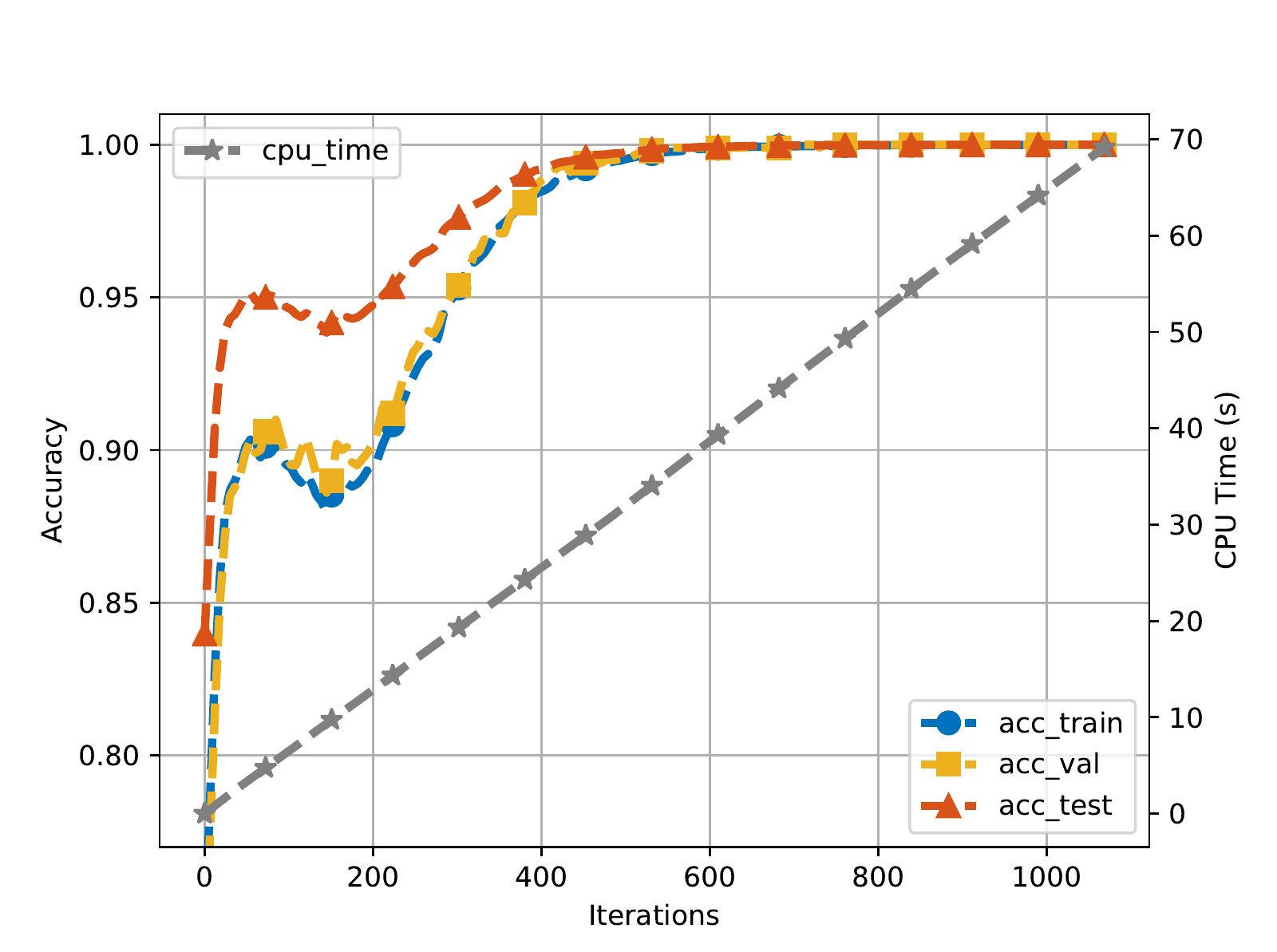}}
	\subfigure[Pre-training ($\delta$=50\%).\label{Fig: 14bus_50_stage1}]{\includegraphics[width=1.7in]{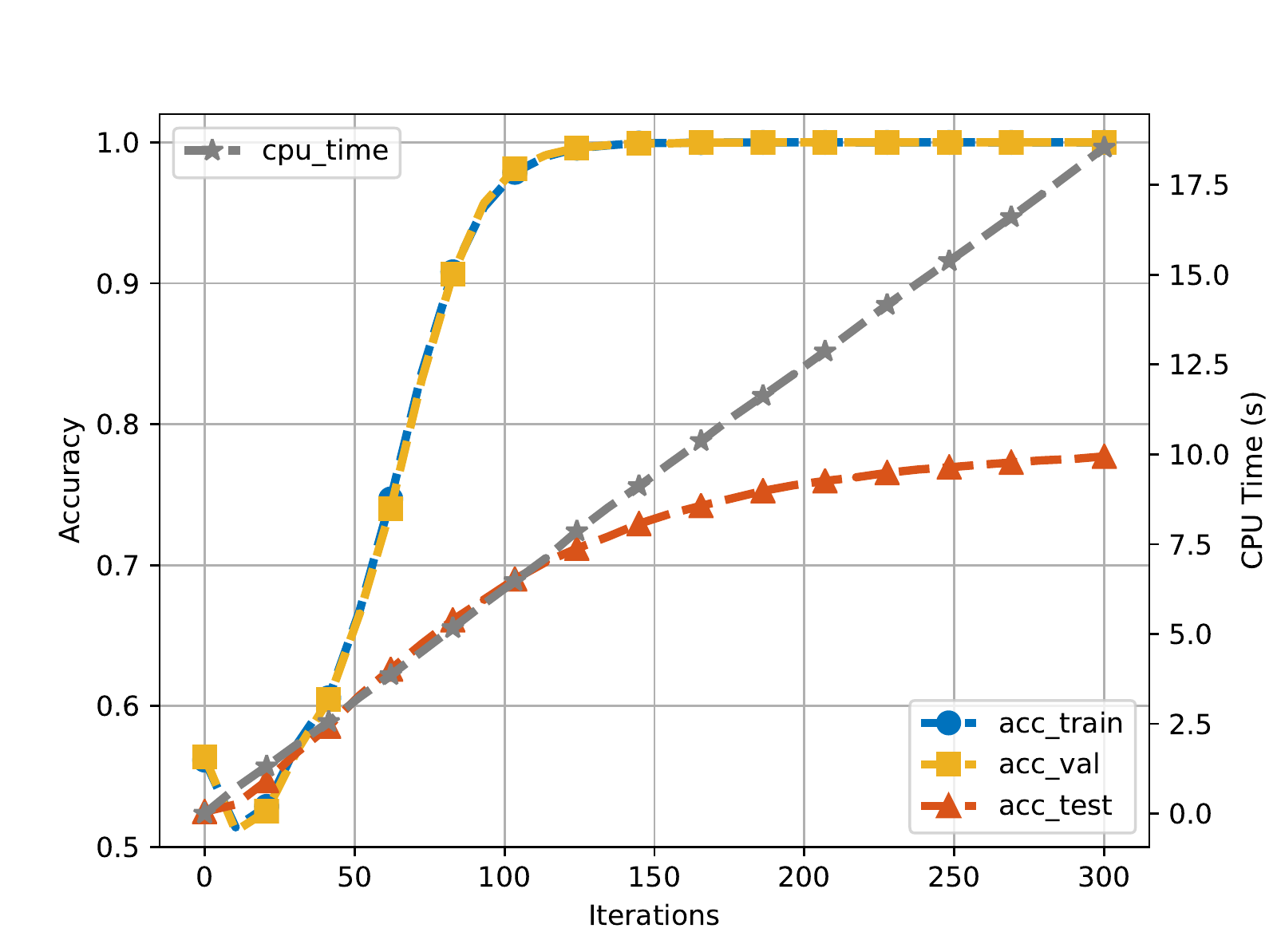}}
	\subfigure[Fine-tuning ($\delta$=50\%).\label{Fig: 14bus_50_stage2}]{\includegraphics[width=1.7in]{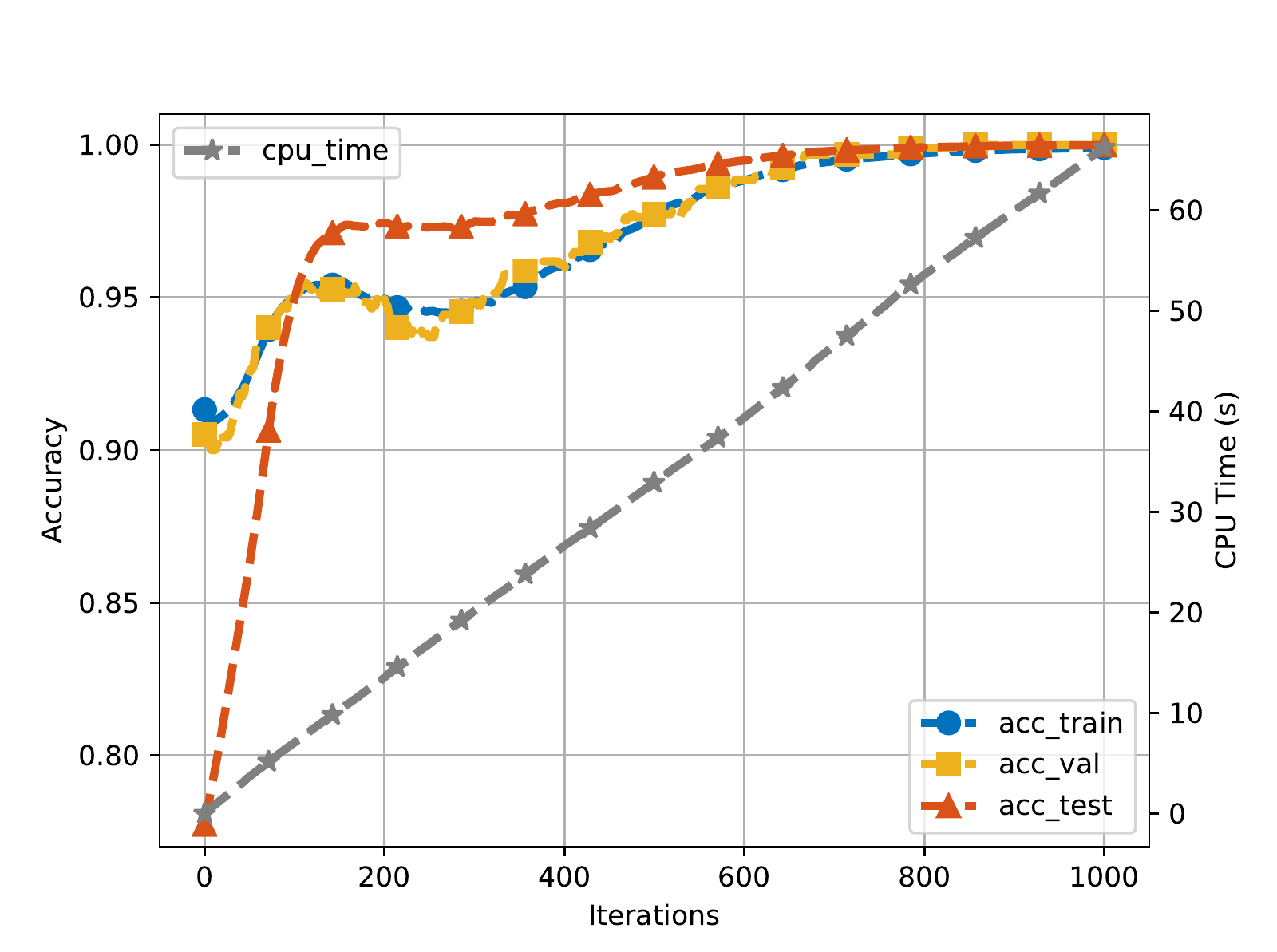}}
	\caption{Convergence performance for the proposed two-stage transfer mechanism on the IEEE 14-bus system with different modeling error levels $\delta$.}
	\vspace{-1.5em}
	\label{Fig: 14bus_convergence performance}
\end{figure}

\begin{figure}[!t]
	\centering
	\subfigure[Pre-training ($\delta$=10\%).\label{Fig: 118bus_10_stage1}]{\includegraphics[width=1.7in]{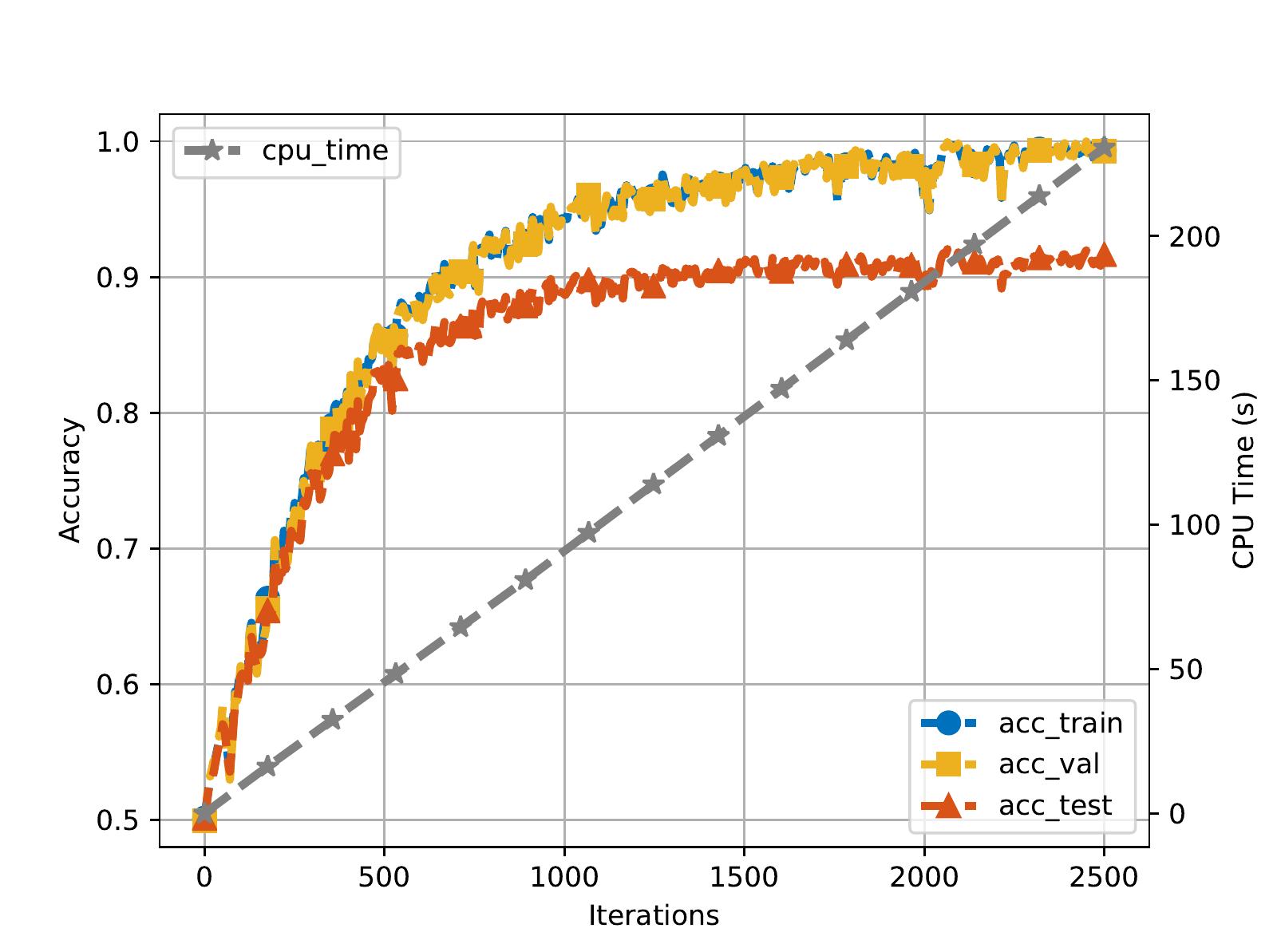}}
	\subfigure[Fine-tuning ($\delta$=10\%).\label{Fig: 118bus_10_stage2}]{\includegraphics[width=1.7in]{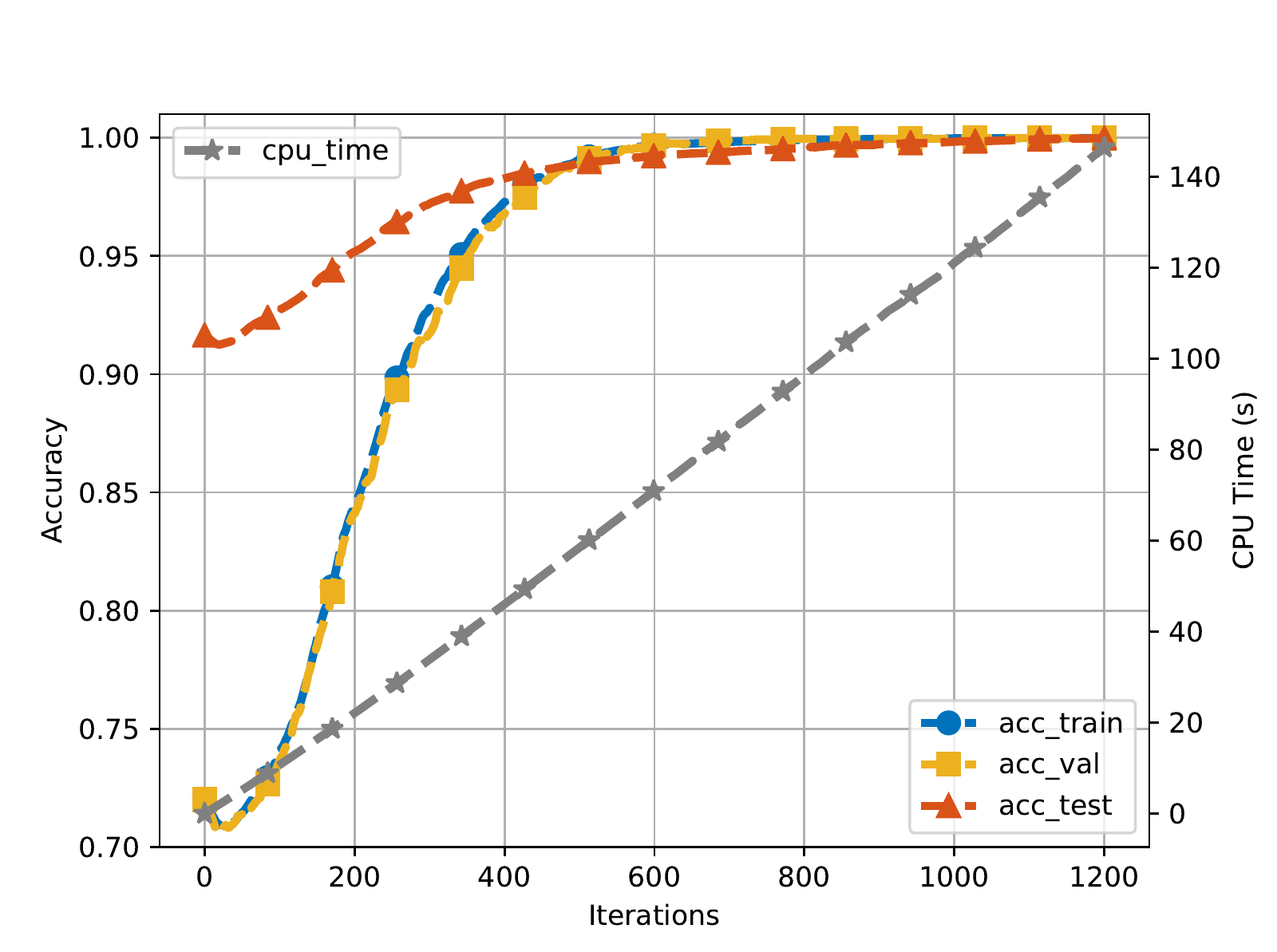}}
	\subfigure[Pre-training ($\delta$=30\%).\label{Fig: 118bus_30_stage1}]{\includegraphics[width=1.7in]{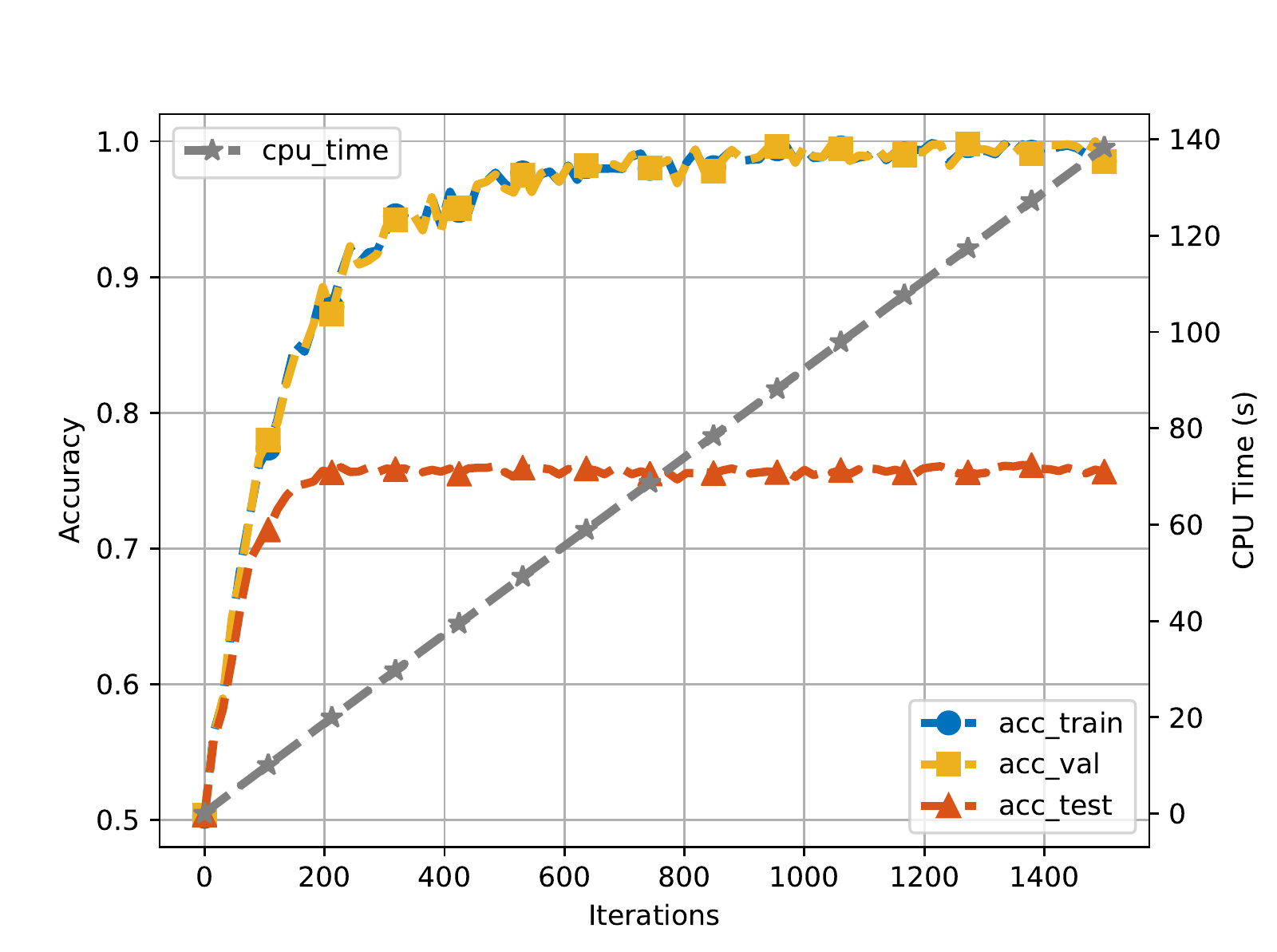}}
	\subfigure[Fine-tuning ($\delta$=30\%).\label{Fig: 118bus_30_stage2}]{\includegraphics[width=1.7in]{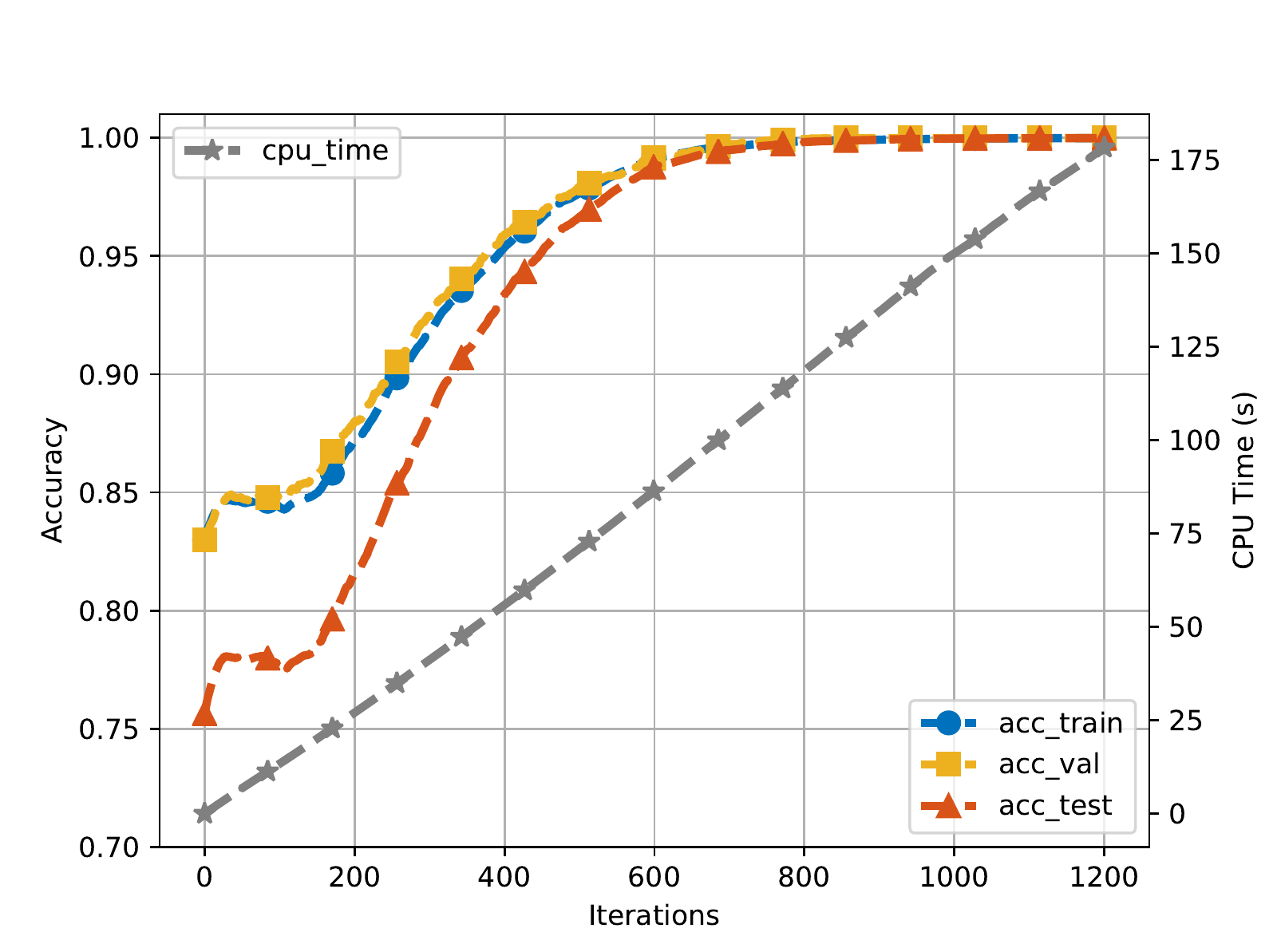}}
	\subfigure[Pre-training ($\delta$=50\%).\label{Fig: 118bus_50_stage1}]{\includegraphics[width=1.7in]{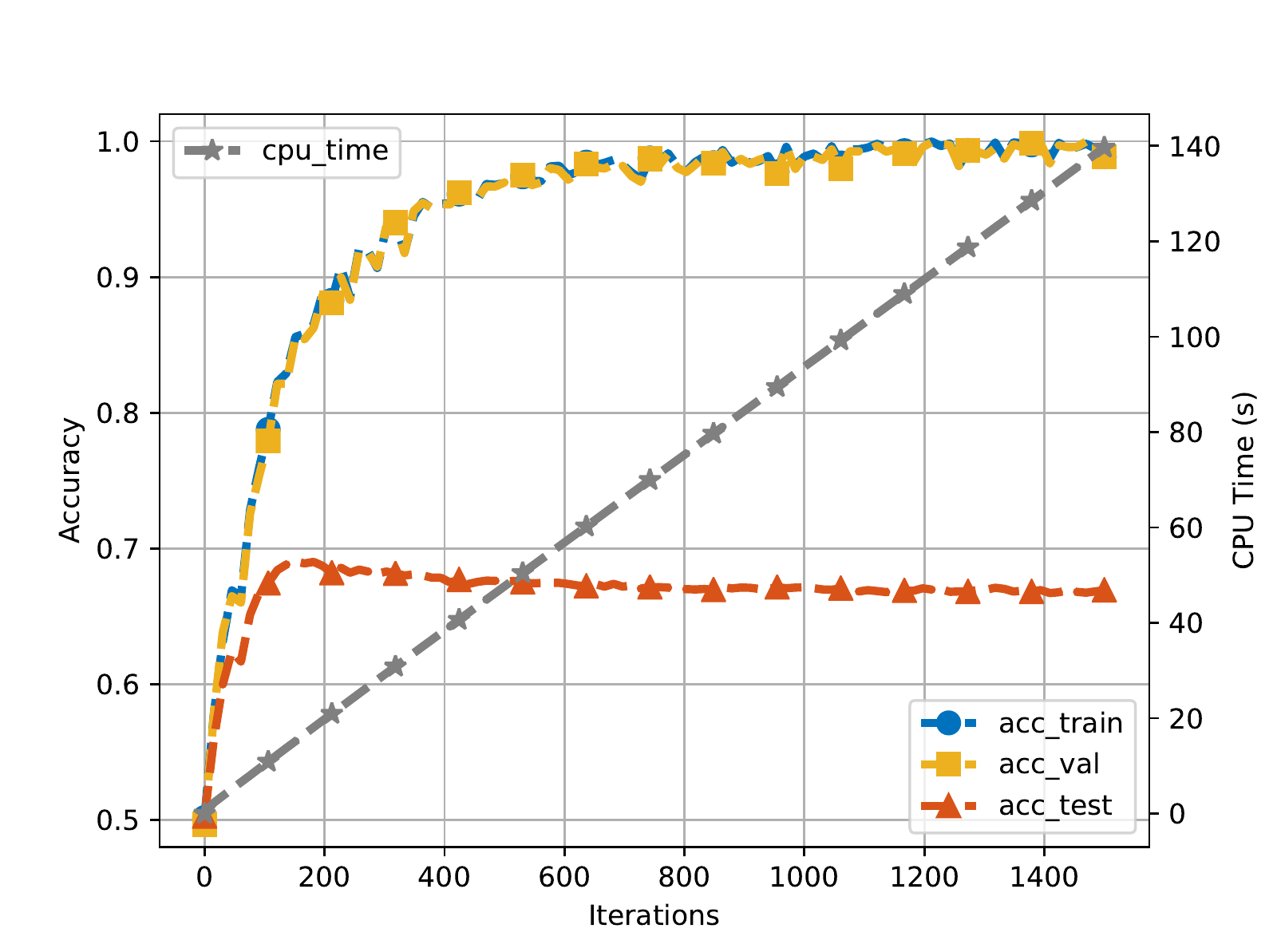}}
	\subfigure[Fine-tuning ($\delta$=50\%).\label{Fig: 118bus_50_stage2}]{\includegraphics[width=1.7in]{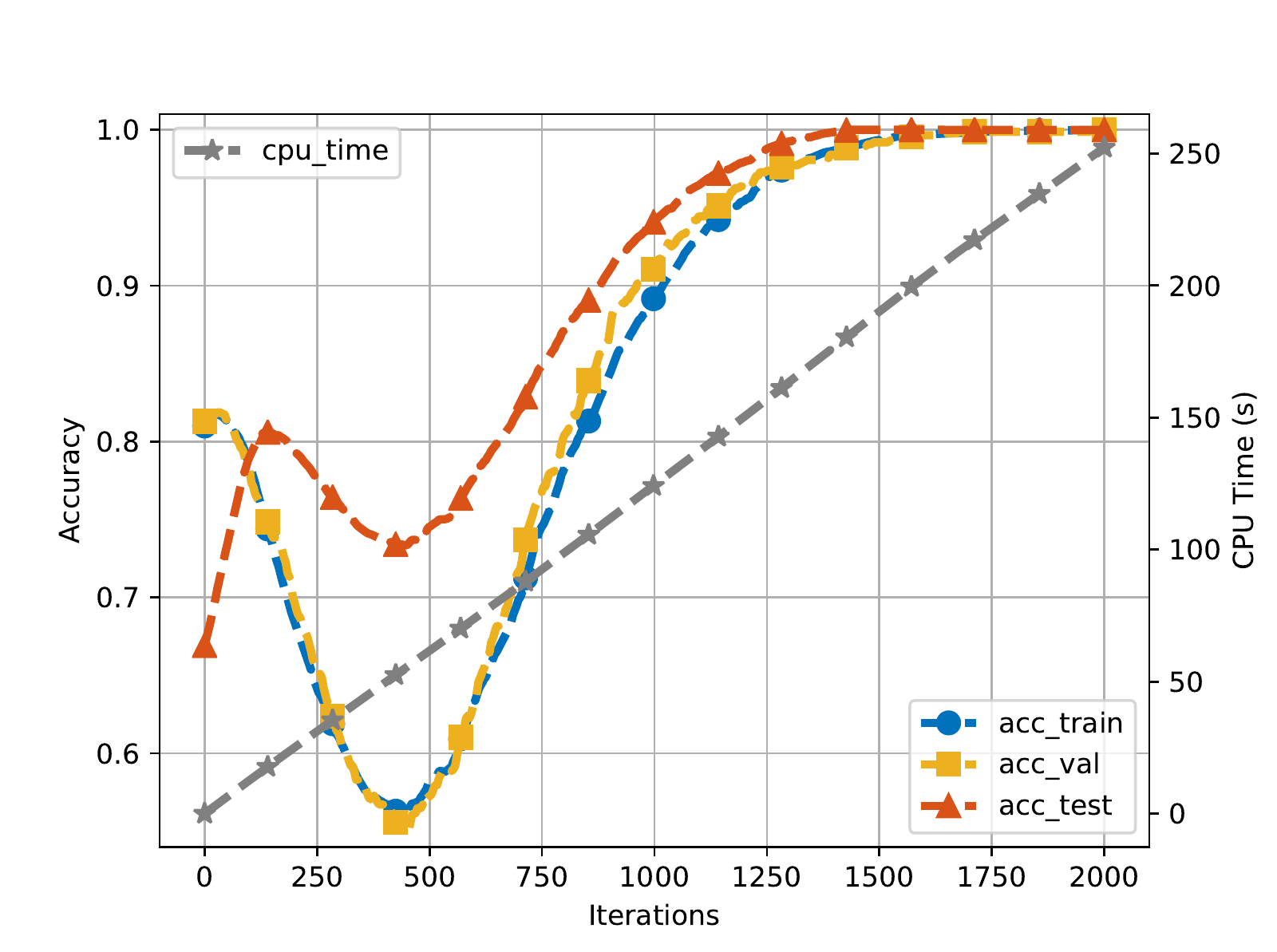}}
	\caption{Convergence performance for the proposed two-stage transfer mechanism on the IEEE 118-bus system with different modeling error levels $\delta$.}
	\vspace{-1.5em}
	\label{Fig: 118bus_convergence performance}
\end{figure}    
\subsection{Detection Performance}\label{Section: Detection Performance}
\subsubsection{Network Transfer}\label{Section: Network Transfer}
The DNN built in our work contains one input layer, three hidden layers, and one output layer, where each hidden layer contains 200 hidden nodes. The outputs of the first $\mathcal{J}=3$ layers are fed into the MMD term. The batch size in two stages is selected as 1e3. The learning rates in the pre-training stage and the fine-tuning stage are set to 1e-3 and 1e-5, respectively. For the involved trade-off hyper-parameters utilized in the pre-training stage, $\lambda$ and $\mu$ are set to 1e-2 and 5e2, respectively.

Then we should mention that a total of 8 scenarios (target domains) with different levels of modeling errors are considered, i.e., $\delta=$ 0\%, 2\%, 5\%, 10\%, 20\%, 30\%, 40\%, 50\%, respectively. Thus, different $\bm{D}_T$ and $\bm{D}_T^*$ are also generated correspondingly. We believe that 50\% covers almost all the system modeling errors that may exist in the real world, hence it can be regarded as the most extreme target domain.  

The two-stage convergence curves of the proposed transfer method on the IEEE 14-bus Case are shown in Fig. \ref{Fig: 14bus_convergence performance} in terms of its ACCs on the training set, validation set, and test set (i.e., $\bm{D}_T^*$) with the system line parameters modeling errors fixed to $\delta=$10\%, 30\%, 50\%, respectively. Let's first look at the scenario $\delta=$10\%. As shown in Fig. \ref{Fig: 14bus_10_stage1}, in the pre-training stage, the ACCs on the training set and the validation set quickly converge to 100\% within 200 iterations, whereas the ACC on $\bm{D}_T^*$ finally converges to nearly 98\%. It is not a surprising result since that, in this stage, only $\bm{D}_S$ is utilized in the supervising part while $\bm{D}_T$ is employed to minimize the joint distribution matching between domains. As shown in Fig. \ref{Fig: 14bus_10_stage2}, in the fine-tuning stage, the ACC on $\bm{D}_T^*$ converges within 800 iterations. The saved DNN model finally shows 99.9913\% ACC on $\bm{D}_T^*$. Also, as shown in Fig. \ref{Fig: 14bus_10_stage1} and \ref{Fig: 14bus_10_stage2}, the CPU time consumed by these two stages are, respectively, 40.22s and 48.68s, which can be considered fairly cheap due to the training acceleration brought by BN and mini-batch methodology. Similar analysis is also applicable to the scenarios $\delta=$30\%, and $\delta=$50\%.

The two-stage convergence curves of the proposed DTL-based method on the IEEE 118-bus Case are shown in Fig. \ref{Fig: 118bus_convergence performance}. The observed convergence performances are similar to those in the IEEE 14-bus Case, thus the involved analysis is omitted here. 
\begin{table*}[!t]
	\centering
	\renewcommand\arraystretch{1.2}
	\tiny
	\caption{Stealthy FDIA Detection Accuracy (\%) and Missing Alarm Rate (\%) Comparison on the IEEE 14-bus System Case with Different Modeling Error Levels}
	\begin{tabular}{cccccccccccccccccc}
		\toprule
		\multirow{1}{*}{Type}&\multirow{1}{*}{Algorithm} &\multicolumn{16}{c}{modeling error level $\delta$} \\
		&&\multicolumn{2}{c}{0\%}&\multicolumn{2}{c}{2\%}&\multicolumn{2}{c}{5\%}&\multicolumn{2}{c}{10\%}&\multicolumn{2}{c}{20\%}&\multicolumn{2}{c}{30\%}&\multicolumn{2}{c}{40\%}&\multicolumn{2}{c}{50\%}\\
		&&ACC&MAR&ACC&MAR&ACC&MAR&ACC&MAR&ACC&MAR&ACC&MAR&ACC&MAR&ACC&MAR\\
		\hline
		\multirow{2}{*}{Model-based}&BDD \cite{liu2011false} &50.00&100.00 &50.00&0.00 &50.00&0.00 &50.00&0.00 &50.00&100.00 &50.00&0.00 &50.00&0.00 &50.00&0.00\\
		&D-FACTS\cite{liu2018reactance} &100.00&0.00 &50.00&0.00 &50.00&0.00 &50.00&0.00  &50.00&0.00 &50.00&0.00 &50.00&0.00 &50.00&0.00\\
		\hline
		\multirow{6}{*}{Data-driven}&SVM &99.86&0.17 &99.82&0.15 &99.80&0.23 &99.42&0.57 &96.92&2.45 &90.83&6.45 &84.28&10.35 &76.90&13.66\\
		&KNN&96.88&3.86 &96.79&4.05 &96.52&4.60 &95.54&5.70 &91.72&9.97 &87.04&15.29 &82.99&21.17 &77.41&26.96\\
		&LR &99.82&0.32 &99.81&0.37 &99.78&0.42 &99.50&0.78 &97.85&3.00 &94.66&8.12 &90.06&14.68 & 85.10&22.28\\
		&RF &98.65&0.14 &95.99&2.93 &95.46&3.34 &94.39&3.75 &88.45&5.79 &81.62&8.44 &77.69&11.58 &71.84&15.24\\
		&GNB&98.33&0.00 &95.41&3.87 &95.09&5.23 &95.04&5.38 &90.53&6.08 &84.57&7.10 &77.46&7.68 &71.03&10.70\\
		&DNN-B&100.00&0.00 &100.00&0.00 &100.00&0.00 &99.50&0.12 &93.73&1.41 &85.72&4.52 &80.12&8.45 &77.74&14.48\\
		&\textbf{Proposed} &\textbf{100.00}&\textbf{0.00} &\textbf{100.00}&\textbf{0.00} &\textbf{100.00}&\textbf{0.00} &\textbf{99.99}&\textbf{0.01} &\textbf{99.99}&\textbf{0.01} &\textbf{99.99}&\textbf{0.01} &\textbf{99.99}&\textbf{0.01} &\textbf{99.99}&\textbf{0.01}\\
		\bottomrule
	\end{tabular}
	\label{Table: 14bus_Detection Accuracy of Different Methods}
\end{table*}

\begin{table*}[!t]
	\centering
	\renewcommand\arraystretch{1.2}
	\tiny
	\caption{Stealthy FDIA Detection Accuracy (\%) and Missing Alarm Rate (\%) Comparison on the IEEE 118-bus System Case with Different Modeling Error Levels}
	\begin{tabular}{cccccccccccccccccc}
		\toprule
		\multirow{1}{*}{Type}&\multirow{1}{*}{Algorithm} &\multicolumn{16}{c}{modeling error level $\delta$} \\
		&&\multicolumn{2}{c}{0\%}&\multicolumn{2}{c}{2\%}&\multicolumn{2}{c}{5\%}&\multicolumn{2}{c}{10\%}&\multicolumn{2}{c}{20\%}&\multicolumn{2}{c}{30\%}&\multicolumn{2}{c}{40\%}&\multicolumn{2}{c}{50\%}\\
		&&ACC&MAR&ACC&MAR&ACC&MAR&ACC&MAR&ACC&MAR&ACC&MAR&ACC&MAR&ACC&MAR\\
		\hline
		\multirow{2}{*}{Model-based}&BDD \cite{liu2011false} &50.00&100.00 &50.00&0.00 &50.00&0.00 &50.00&0.00 &50.00&100.00 &50.00&0.00 &50.00&0.00 &50.00&0.00\\
		&D-FACTS\cite{liu2018reactance} &100.00&0.00 &50.00&0.00 &50.00&0.00 &50.00&0.00  &50.00&0.00 &50.00&0.00 &50.00&0.00 &50.00&0.00\\
		\hline
		\multirow{6}{*}{Data-driven}&SVM &68.37&29.74 &66.71&29.54 &66.12&30.66 &66.05&31.98 &64.83&36.47 &63.45&44.43 &61.53&53.84 &58.66&63.00\\
		&KNN&56.09&34.36 &55.40&34.22 &55.42&35.27 &55.77&34.86 &55.01&34.10 &55.07&35.30 &54.51&34.67 &54.48&33.19\\
		&LR &93.28&3.70 &93.12&3.90 &91.88&5.51 &88.56&10.62 &81.01&21.41 &75.71&24.19 &71.69&27.03 & 69.17&30.12\\
		&RF &57.56&35.10 &55.64&35.17 &54.45&35.07 &55.22&34.63 &54.69&35.32 &54.53&35.20 &54.38&35.54 &53.91&35.04\\
		&GNB&70.70&46.35 &59.21&66.80 &55.89&49.38 &55.66&46.23 &55.64&44.76 &55.72&43.95 &54.77&44.58 &55.22&44.14\\
		&DNN-B&100.00&0.00 &99.95&0.03 &98.90&1.31 &91.65&10.34 &81.73&25.90 &75.64&28.39 &71.40&34.36 &66.92&37.07\\
		&\textbf{Proposed} &\textbf{100.00}&\textbf{0.00} &\textbf{100.00}&\textbf{0.00} &\textbf{99.99}&\textbf{0.01} &\textbf{99.99}&\textbf{0.01} &\textbf{99.99}&\textbf{0.01} &\textbf{99.99}&\textbf{0.01} &\textbf{99.99}&\textbf{0.01} &\textbf{99.99}&\textbf{0.01}\\
		\bottomrule
	\end{tabular}
	\label{Table: 118bus_Detection Accuracy of Different Methods}
\end{table*}

\subsubsection{Comparison Baselines}
In order to verify the effectiveness of the proposed transfer mechanism, some existing baselines ignoring the involved modeling errors are conducted for comparison. For the model-based approaches, the conventional state estimation based BDD algorithm is first incorporated into our baselines to verify the stealthiness of the generated FDIA. Then a popular D-FACTS devices based method developed from \cite{liu2018reactance} is implemented, where the residual threshold is set to 0.05. 

In addition, the involved data-driven methods contain
\begin{enumerate}
	\item[1)] Logistic Regression (LR); 
	
	\item[2)] Support Vector Machine (SVM); 
	
	\item[3)] K-Nearest Neighbors (KNN);
	
	\item[4)] Random Forest (RF);
	
	\item[5)] Gaussian Naive Bayes (GNB);
	
	\item[6)] Deep Neural Network (DNN-B).	
\end{enumerate}

Note that for fair comparisons, the involved hyper-parameters in these baselines are selected from a set while the best performances are recorded. Specifically, the trade-off parameter for LR is selected by searching \{1e-3,1e-2,1e-1,1,1e1,1e2\}. SVM is implemented with the Gaussian kernel and the punishment factor is taken from \{1e-1,2e-1,$\ldots$,1,2,4,8,16,32,64\}. The number of nearest neighbors for KNN is selected from \{1,2,5,10,50\}. The size of the estimators is set to 8 in RF. DNN-B denotes a DNN-based classifier that has the same network structure as the one implemented in our approach. 1)-6) are trained with $\bm{D}_S$ and tested via $\bm{D}_T^*$.

\subsubsection{Evaluation Metrics}
Two typical metrics for binary classification tasks are employed in this paper:
\begin{enumerate}
	\item[$\bigcdot$] Accuracy (ACC): $ACC = \frac{TP+TN}{TP+TN+FP+FN}$ 

	\item[$\bigcdot$] Missing Alarm Rate (MAR): $MAR = \frac{FN}{TP+FN}$ 
\end{enumerate}
where TP, TN, FP, FN denote the number of true positives, true negatives, false positives, and false negatives, respectively. The MAR is also known as false negative rate.   

\subsubsection{Detection Results}\label{Section: Detection Results}   
Table \ref{Table: 14bus_Detection Accuracy of Different Methods} and Table \ref{Table: 118bus_Detection Accuracy of Different Methods}, respectively, report the comparison results between the proposed mechanism and other baselines in terms of ACC and MAR on the IEEE 14-bus and 118-bus test sets. Note that for each baseline, 100 independent trials are conducted and the average ACC is reported. Firstly, we can observe that two model-based FDIA methods are very sensitive to modeling errors since the presence of modeling errors causes the obtained residual to exceed the preset threshold and thus both normal samples and attack samples are detected as attacks. Further, we easily find that the ACC performances of data-driven baselines decline with increasing modeling errors $\delta$, whereas the proposed mechanism outperforms those baselines in all scenarios, especially the scenario with $\delta$ being 50\%. Note that the ACCs of the proposed mechanism exceed 99\% in all scenarios, which can be considered very satisfactory in comparison with other baselines. It is worth mentioning that the proposed approach fully surpasses DNN-B, revealing a better FDIA detection performance of the proposed two-stage DTL-based method than conventional deep learning-based approaches, which echoes our statements in Remark 3.      
\begin{table}[!t]
	\centering
	\renewcommand\arraystretch{1.2}
	\tiny
	\caption{Stealthy FDIA Detection Accuracy (\%) Comparison on the IEEE 14-bus System Case with Unknown Measurement Noise in Target Domain}
	\begin{tabular}{ccccccccc}
		\toprule
		\multirow{1}{*}{Type}&\multirow{1}{*}{Algorithm} &\multicolumn{7}{c}{Measurement noise $\sigma$ in source domain $H$} \\
		&&0\%&1\%&2\%&3\%&5\%&7\%&10\%\\
		\hline
		\multirow{2}{*}{Model-based}&BDD \cite{liu2011false} &50.00&50.00 &50.00&50.00 &50.00 &50.00 &50.00\\
		&D-FACTS\cite{liu2018reactance} &50.00&50.00 &50.00&50.00 &50.00 &50.00 &50.00\\
		\hline
		\multirow{6}{*}{Data-driven}
		&SVM &77.24&76.90 &73.45 &72.29 &71.72 &70.40 &68.36\\
		&KNN &77.97 &77.41 &76.28 &75.51 &74.60 &72.53 &70.24\\
		&LR  &85.59 &85.10 &82.78 &82.60 &82.42 &81.60 &79.51\\
		&RF  &72.21 &71.84 &70.95 &72.01 &72.38 &72.21 &71.55\\
		&GNB &70.48 &71.03 &72.07 &72.56 &73.24 &75.86 &76.47\\
		&DNN-B&77.56&77.74 &77.03 &75.10 &74.11 &55.24 &51.59\\
		&\textbf{Proposed} &\textbf{99.99}&\textbf{99.99} &\textbf{99.99}&\textbf{99.99} &\textbf{99.99} &\textbf{95.76} &\textbf{56.06}\\
		\bottomrule
	\end{tabular}
	\label{Table: 14bus_Impact of Measurement Noise}
\end{table}
\begin{table}[!t]
	\centering
	\renewcommand\arraystretch{1.2}
	\tiny
	\caption{Stealthy FDIA Detection Accuracy (\%) Comparison on the IEEE 118-bus System Case with Unknown Measurement Noise in Target Domain}
	\begin{tabular}{ccccccccc}
		\toprule
		\multirow{1}{*}{Type}&\multirow{1}{*}{Algorithm} &\multicolumn{7}{c}{Measurement noise $\sigma$ in source domain $H$} \\
		&&0\%&1\%&2\%&3\%&5\%&7\%&10\%\\
		\hline
		\multirow{2}{*}{Model-based}&BDD \cite{liu2011false} &50.00&50.00 &50.00&50.00 &50.00 &50.00 &50.00\\
		&D-FACTS\cite{liu2018reactance} &50.00&50.00 &50.00&50.00 &50.00 &50.00 &50.00\\
		\hline
		\multirow{6}{*}{Data-driven}
		&SVM &58.10 &58.66 &58.51 &57.95 &56.06 &55.32 &53.36\\
		&KNN &54.42 &54.48 &54.33 &54.00 &53.27 &52.85 &51.73\\
		&LR  &69.90 &69.17 &66.37 &65.78 &65.02 &63.29 &62.45\\
		&RF  &53.99 &53.91 &53.75 &52.19 &53.54 &52.86 &52.56\\
		&GNB &54.31 &55.22 &54.02 &55.14 &56.41 &56.09 &56.77\\
		&DNN-B&66.24&66.92 &66.10 &65.80 &64.99 &58.43 &54.61\\
		&\textbf{Proposed} &\textbf{99.99}&\textbf{99.99} &\textbf{99.99}&\textbf{99.99} &\textbf{99.99} &\textbf{92.14} &\textbf{62.01}\\
		\bottomrule
	\end{tabular}
	\label{Table: 118bus_Impact of Measurement Noise}
\end{table}

\subsection{The Impact of $\sigma$}
In this subsection, we investigate the impact of inaccurate estimation of measurement noise on the performance of the proposed approach. Specifically, we set the measurement noise in $H^*$ to $\sigma$=1\%, while a series of $H$ with $\sigma$=0\%, 1\%, 2\%, 3\%, 5\%, 7\%, and 10\% are respectively considered. The modeling errors is fixed to $\delta$=50\%. The detection results are shown in Table \ref{Table: 14bus_Impact of Measurement Noise} and Table \ref{Table: 118bus_Impact of Measurement Noise}. We can observe that when the $\sigma$ in $H$ does not exceed 5\%, the performance of the proposed method will not be significantly affected. When the $\sigma$ in $H$ exceeds 7\%, the ACC of the proposed method drops significantly. Similar behavior can also be observed on DNN-B. The rationale behind this case is that, compared with other baselines, the proposed method and DNN-B are both deep neural networks whose performances are prone to be affected by large noise due to their parameterized structure. However, it is worth pointing out that $\sigma$ over 7\% is an extreme case since the one in real world is usually ensured to fairly small (e.g., 1\%) so as to guarantee the performance of conventional state estimation \cite{yu2018online}. Therefore, as a summary of this subsection, it is advocated that the $\sigma$ in $H$ should be estimated as close as possible to the one in $H^*$, while the proposed method shows a considerably satisfactory error tolerance.

\section{Conclusion}\label{Section: Conclusion}
In this paper, we propose a deep transfer learning-based approach to detect the FDIA in smart grids considering system line parameters modeling errors. The involved transfer strategy contains two stages. In the first one, three refined terms are simultaneously optimized, which aims at improving the detection performance of the built DNN on the simulation (source domain) dataset while minimizing the joint distribution matching between domains. The second stage uses the real (target domain) data to fine-tune the trained deep neural network. A significant advantage of the proposed mechanism is that it can trace the system modeling errors online by only collecting real normal data, which is considerably practical in real smart grids.

Our future research may include two parts. On one hand, how to select the hyper-parameters contained in the proposed mechanism is an interesting but challenging topic. On the other hand, how to implement a simpler transfer while maintaining excellent transfer performance has also drawn our attention.

\bibliographystyle{IEEEtran}
\bibliography{ref}
\end{spacing}
\end{document}